\documentclass[11pt,oneside,fleqn,a4paper]{article}

\usepackage[ansinew]{inputenc}
\usepackage{hyperref}
\usepackage[mathscr]{eucal}
\usepackage{amsmath,amssymb,amsthm}
\usepackage{graphicx}
\usepackage{color}

\allowdisplaybreaks

\hypersetup{colorlinks, linkcolor=blue, citecolor=blue, urlcolor=blue}
\makeatletter
\long\def\@makecaption#1#2{%
  \vskip\abovecaptionskip\footnotesize
  \sbox\@tempboxa{#1. #2}%
  \ifdim \wd\@tempboxa >\hsize
    #1. #2\par
  \else
    \global \@minipagefalse
    \hb@xt@\hsize{\hfil\box\@tempboxa\hfil}%
  \fi
  \vskip\belowcaptionskip}
\makeatother

\newcommand{\dd}[2]{\frac{\mathrm{d} #1}{\mathrm{d} #2}}
\newcommand{\pdl}[2]{\frac{\partial #1}{\partial #2}}

\newcommand{\ddl}[2]{\frac{\delta #1}{\delta #2}}
\newcommand{\ddd}{\mathrm{d}}
\newcommand{\p}{\partial}

\newcommand{\todo}[1][\null]{\ensuremath{\clubsuit}}

\newcommand{\noprint}[1]{}

{\theoremstyle{definition}

\newtheorem{remark}{Remark}
\newtheorem*{remark*}{Remark}
}

\newcommand{\checked}[1][\null]{\ensuremath{\boldsymbol{\surd}}}

\newcommand{\vv}{\mathbf{v}}

\newcommand{\kk}{\mathbf{k}}

\newcommand{\ve}{\varepsilon}

\newcommand{\nn}{\nabla}

\begin{document}

\par\noindent {\LARGE\bf
Invariant discretization schemes\texorpdfstring{\\}{ }for the shallow-water equations
\par}

{\vspace{4mm}\par\noindent {\large Alexander Bihlo$^\dag$ and Roman O.\ Popovych$^\ddag$
} \par\vspace{2mm}\par}

{\vspace{2mm}\par\noindent {\it
$^{\dag}$Centre de recherches math\'{e}matiques, Universit\'{e} de Montr\'{e}al, C.P.\ 6128, succ.\ Centre-ville,\\$\phantom{^\dag}$Montr\'{e}al (QC) H3C 3J7, Canada\\
}}
{\noindent \vspace{2mm}{\it
$\phantom{^\dag}$\textup{E-mail}: bihlo@crm.umontreal.ca
}\par}

{\vspace{2mm}\par\noindent {\it
$^{\ddag}$Faculty of Mathematics, University of Vienna, Nordbergstra{\ss}e 15, A-1090 Vienna, Austria\\
$\phantom{^\dag}$Institute of Mathematics of NAS of Ukraine, 3 Tereshchenkivska Str., 01601 Kyiv, Ukraine
}}

{\noindent \vspace{2mm}{\it
$\phantom{^\dag}$\textup{E-mail}: rop@imath.kiev.ua
}\par}

\vspace{6mm}\par\noindent\hspace*{8mm}\parbox{144mm}{\small\looseness=-1
Invariant discretization schemes are derived for the one- and two-dimensional shallow-water equations with periodic boundary conditions. While originally designed for constructing invariant finite difference schemes, we extend the usage of difference invariants to allow constructing of invariant finite volume methods as well. It is found that the classical invariant schemes converge to the Lagrangian formulation of the shallow-water equations. These schemes require to redistribute the grid points according to the physical fluid velocity, i.e., the mesh cannot remain fixed in the course of the numerical integration. Invariant Eulerian discretization schemes are proposed for the shallow-water equations in computational coordinates. Instead of using the fluid velocity as the grid velocity, an invariant moving mesh generator is invoked in order to determine the location of the grid points at the subsequent time level. The numerical conservation of energy, mass and momentum is evaluated for both the invariant and non-invariant schemes.
}\par\vspace{4mm}

\noprint{
Key words: invariant discretization schemes, Lie symmetries, shallow-water equations, difference invariants, adaptive mesh refinement
MSC: 76M60, 76M20, 65M50
76-XX   Fluid mechanics For general continuum mechanics, see 74Axx, or other parts of 74-XX
 76Mxx  Basic methods in fluid mechanics [See also 65-XX]
  76M20   Finite difference methods
  76M60   Symmetry analysis, Lie group and algebra methods
86-XX   Geophysics
  86-08   Computational methods
65-XX   Numerical analysis
 65Mxx  Partial differential equations, initial value and time-dependent initial-boundary value problems
  65M50   Mesh generation and refinement
}




\section{Introduction}

Discretization schemes that preserve characteristic properties of systems of differential equations have received increasing attention over the past years and led to the development of the field of \emph{geometric numerical integration}. The principal motivation for this approach is that controlling the local discretization error, as is done in most of the classical numerical methods, can fail to capture essential qualitative features of the underlying problem, which might be equally important in order to obtain reasonable integration results. Such features can include, but are not necessarily limited to, conservation laws, point symmetries, Hamiltonian structure, conservation of phase-space volume and asymptotic characteristics. Various geometric numerical integration schemes have been developed that capture these properties in the course of discretization, such as conservation laws and the Hamiltonian structure~\cite{brid06Ay,leim04Ay,reic99Ay}, Lie symmetries~\cite{baki97Ay,chha10Ay,chha11Ay,doro11Ay,kim08Ay,vali05Ay} and phase-space volume~\cite{egge96Ay,somm10Ay}.

In the present paper, we aim to concentrate on the problem of deriving discretization schemes with symmetry properties by developing appropriate finite difference and finite volume schemes for the shallow-water equations. In particular, we are concerned with the problem of finding discretization schemes that are invariant under the maximal Lie invariance group admitted by the shallow-water equations with (double) periodic boundary conditions. Choosing the shallow-water equations for such an investigation can be motivated because they constitute a prominent, simple, yet fully-nonlinear model of fluid mechanics exhibiting various features of the original set of governing equations of hydrodynamics, such as the simultaneous occurrence of both fast (divergent) and slow (vortical) waves and the existence of conservation laws, symmetries and a Hamiltonian form. Moreover, the shallow-water equations always served as an important intermediate model to test new numerical schemes \cite{arak81Ay,fran04Ay,reic99Ay,ring02Ay,salm07Ay,somm09Ay}.

Similarly to conservation laws, symmetries have important implications on the solutions of differential equations. When simulating the dynamics of a classical mechanical system in a constantly moving coordinate system, it should be a clear desire that the numerical model to be used for that problem is Galilean invariant as otherwise physical laws can be violated. It is also well known \cite{budd09Ay,budd99Ay,huan10By} that the shape of a solution of a system of differential equations near a blow-up point can tend to a group-invariant solution of this system. Often group-invariant solutions well capture so-called intermediate asymptotic behavior of the solutions after a sufficiently long period of evolution. For the simulation of invariant solutions, symmetry-preserving discretization schemes can give better numerical results than standard schemes that do not preserve the geometry of differential equations.

The design of invariant discretization schemes for evolution equations in general requires the explicit treatment of meshes that are not time-space orthogonal, i.e., time-adaptive grids. Such grids pose several challenges from the numerical point of view that have not been well investigated in the field of invariant numerical schemes up to now. On the other hand, meshes that adapt according to the development of the numerical solution are an extensively investigated subject in the field of numerical mathematics, see, e.g., \cite{huan10By,thom85By}. The question not explicitly answered so far is whether the problem of finding discretization schemes with symmetry properties can be embedded into the study of adaptive numerical schemes in the multidimensional case. In the present paper, we discuss a possible answer to that problem, exemplified by the shallow-water equations.

The outline of the paper is as follows: Properties of the shallow-water equations are discussed in Section~\ref{sec:SymmetriesShallowWater}. Section~\ref{sec:ConstructionOfInvariantSchemes} is devoted to a review of common techniques that allow one to construct invariant finite difference schemes. In Section~\ref{sec:OneDimensionalShallowWater} we derive invariant discretization schemes for the one-dimensional shallow-water equations. This is done both by using the Lagrangian description of the shallow-water equations and by setting up an invariant grid generator for Eulerian schemes in computational coordinates. In Section~\ref{sec:TwoDimensionalShallowWater} we discuss strategies for the design of invariant numerical models in higher dimensions and illustrate them with the two-dimensional shallow-water equations. Again, both Lagrangian schemes in physical coordinates and Eulerian schemes in computational coordinates with an invariant grid generator are introduced. For the first scheme we use an invariant finite volume discretization, while the second scheme is based on finite differences. A summary and concluding remarks can be found in Section~\ref{sec:Conclusion}.

\section{Symmetries of the shallow-water equations}\label{sec:SymmetriesShallowWater}

The nondimensionalized system of shallow-water equations in Cartesian coordinates is
\begin{align}\label{eq:TwoDimensionalShallowWater}
\begin{split}
&u_t + uu_x +vu_y + h_x = 0, \\
&v_t + uv_x + vv_y + h_y = 0, \\
&h_t + uh_x + vh_y + h(u_x+v_y) = 0,
\end{split}
\end{align}
where $\vv=(u,v)$ is the fluid velocity in the plane and $h$ is the height of the fluid column over a fixed reference level within the fluid. The bottom topography is assumed to be flat here for simplicity. Treating non-flat topographies would lead to the inclusion of additional source terms in system~\eqref{eq:TwoDimensionalShallowWater}. The shallow-water equations are derived from the Euler equations for an ideal fluid under the following assumptions: the validity of the hydrostatic approximation, constancy of the fluid density and much smaller scale of vertical motions in comparison with horizontal motions~\cite{pedl87Ay}.

The shallow-water equations~\eqref{eq:TwoDimensionalShallowWater} can be represented in Hamiltonian form~\cite{morr80Ay} using
\[
 \{\mathcal F,\mathcal G\} = \int\left(q\kk\cdot\ddl{\mathcal F}{\vv}\times\ddl{\mathcal G}{\vv}-\ddl{\mathcal F}{\vv}\cdot\nn\ddl{\mathcal G}{h}+\ddl{\mathcal G}{\vv}\cdot\nn\ddl{\mathcal F}{h} \right)\ddd A
\]
as a Poisson bracket, where $\mathcal F$ and $\mathcal G$ are functionals of~$\vv$ and~$h$, $q=\zeta/h=(v_x-u_y)/h$ is the potential vorticity, $\kk$ denotes the vertical unit vector, $\ddd A=\ddd x\ddd y$ is the area element, and the integration extends over the domain of the entire fluid. The Hamiltonian for the shallow-water equations is given by the total energy
\[
 \mathcal H = \frac12\int \left(h\vv^2+h^2\right)\ddd A.
\]
Additional conserved quantities are associated with the above noncanonical Poisson bracket. For any function $f$ of the potential vorticity $q$, the integral
\[
 \mathcal C_f = \int hf(q)\,\ddd A
\]
is conserved on solutions of the shallow-water equations. This class of conserved quantities contains the mass $\mathcal M=\mathcal C_1$, the circulation $\mathcal Z = \mathcal C_{q}$ and the potential enstrophy $\mathcal E= \mathcal C_{q^2/2}$.
Two more conserved quantities are the momenta in the $x$- and $y$-directions,
\[\mathcal P_x = \int hu\, \ddd A,\quad
\mathcal P_y = \int hv\, \ddd A.\]

The maximal Lie invariance algebra~$\mathfrak g_2$ of the two-dimensional shallow-water equations~\eqref{eq:TwoDimensionalShallowWater} is nine-dimensional; see, e.g., \cite{ches09Ay,pavl05Ay}. A basis of this algebra consists of the vector fields
\begin{align*}
\begin{split}
    & \p_t,\quad \p_x,\quad \p_y, \quad t\p_x+\p_u,\quad t\p_y+\p_v,\\
    & t\p_t+x\p_x+y\p_y,\quad x\p_x+y\p_y+u\p_u+v\p_v+2h\p_h,\\
    & -y\p_x+x\p_y-v\p_u+u\p_v,\quad t^2\p_t+tx\p_x+ty\p_y + (x-tu)\p_u+(y-tv)\p_v-2th\p_h.
\end{split}
\end{align*}
These vector fields generate one-parameter Lie symmetry groups, which correspond to (i) time translations, (ii)--(iii) space translations, (iv)--(v) Galilean transformations, (vi)--(vii) scalings, (viii) rotations and (ix) inversions in~$t$.

In what follows, we will also use the one-dimensional version of system~\eqref{eq:TwoDimensionalShallowWater}, in which case we set $v=0$ and drop the dependence of $u$ and $h$ on $y$. The resulting system reads
\begin{align}\label{eq:OneDimensionalShallowWater}
\begin{split}
&u_t + uu_x + h_x = 0, \quad
h_t + uh_x + hu_x = 0
\end{split}
\end{align}
and preserves the one-dimensional versions of total energy, mass and momentum,
\[
 \mathcal H = \frac12\int \left(hu^2+h^2\right)\ddd x, \qquad
 \mathcal M = \int h\, \ddd x, \qquad
 \mathcal P = \int hu\, \ddd x.
\]

It is well known that the maximal Lie invariance algebra~$\mathfrak g_1$ of system~\eqref{eq:OneDimensionalShallowWater} is infinite dimensional and spanned by the vector fields
\begin{align*}
\begin{split}
&t\p_t + x\p_x, \quad x\p_x+u\p_u+2h\p_h,\quad  t\p_x + \p_u,\\
&(2x-6tu)\p_t + (6h-3u^2)t\p_x+(u^2+4h)\p_u+4hu\p_h,\quad f(h,u)\p_t+ g(h,u)\p_x,
\end{split}
\end{align*}
where the functions~$f$ and~$g$ run through the set of solutions of the system
\begin{equation}\label{eq:OneDimShallowWaterLinearized}
\begin{split}
&g_h-uf_h+f_u=0,\quad g_u-uf_u+hf_h=0.
\end{split}
\end{equation}
The existence of the latter generator is owed to the possibility of
linearization of system~\eqref{eq:OneDimensionalShallowWater} to system
\eqref{eq:OneDimShallowWaterLinearized} by means of the hodograph
transformation in that $u$ and~$h$ are assumed as the new independent
variables and $f=t$ and $g=x$ are the new unknown functions. The
linearization by the hodograph transformation permuting the pairs of
dependent and independent variables is a general property of homogeneous
first-order systems of partial differential equations in two independent
variables and two unknown functions that are linear in derivatives with
coefficients depending only on the unknown functions. See
also~\cite[p.~154]{hydo00Ay} for the symmetry interpretation of the
linearization of the one-dimensional shallow-water equations.
Note that system~\eqref{eq:OneDimShallowWaterLinearized} is reduced to a single Tricomi equation.
More precisely, excluding~$g$ by cross differentiation, we obtain the equation $f_{uu}=hf_{hh}+2f_h$.
The substitution $\phi=hf$ then leads to the Tricomi equation $\phi_{uu}=h\phi_{hh}$.
Another way to reduce the system is to rewrite the equation $f_{uu}=hf_{hh}+2f_h$ in the form $h^3f_{uu}=h^2(h^2f_h)_h$
and to carry out the transformation $z=1/h$, which yields a similar Tricomi equation, $f_{uu}=z^3f_{zz}$.
Symmetry analysis of such equations was carried out in~\cite{blum87Ay,blum88Ay}.

\section{Construction of invariant numerical discretization schemes}\label{sec:ConstructionOfInvariantSchemes}

\looseness=-1
The problem of constructing discretization schemes that preserve symmetries of the corresponding differential equations was first systematically addressed by Dorodnitsyn and his collaborators~\cite{baki97Ay,budd01Ay,doro11Ay,doro91Ay,doro03Ay,doro00Ay}. As there are an infinite number of possibilities to approximate a differential equation by means of finite differences, one might single out those among all possible difference schemes that inherit symmetries of the original differential equation. Dorodnitsyn's approach can be summarized in the following way: First determine the maximal Lie invariance algebra of the model under consideration. For many classical hydrodynamical problems, this task has already been completed~\cite{ibra95Ay,ovsi82Ay}. Then a discretization stencil has to be chosen.
The generators of Lie symmetries are then prolonged to all points of the stencil. From these prolonged generators, the invariants of the extended group action on the stencil are determined. The final step is then to assemble the obtained invariants into a difference approximation of the original differential equation. By difference approximation it is meant that in the continuous limit, the invariant finite difference scheme reduces to the original differential equation in some coordinates. Each difference approximation consists of a physical difference equation and equations governing the positions of grid points. In the continuous limits, these grid equations often reduce to some trivial identities.

Altogether, this method is a straightforward application of \emph{inverse group classification}, using transformation groups acting on functions defined on a discrete set of points rather than on a continuous space. In the usual inverse group classification one starts with a particular Lie group~$G$ and aims at finding those systems of differential equations that admit~$G$ as a symmetry group. In practice these systems are found by computing differential invariants (i.e., invariants that involve derivatives of dependent variables) of~$G$. Any function of differential invariants is a differential invariant of~$G$ and, subject to some regularity condition, any system of differential equations can be expressed in terms of differential invariants of its maximal Lie invariance group~\cite{olve09Ay}. The Dorodnitsyn method works by selecting the maximal Lie symmetry group of a system of differential equations as the initial Lie group~$G$. By extending the action of~$G$ to the points of the discretization stencil, one is able to compute invariants of the extended action, i.e., difference invariants of~$G$. As in the continuous case, any function of difference invariants is a difference invariant. Constructing a difference approximation of a system of differential equations using difference invariants therefore leads to a symmetry-preserving discretization scheme.

A common feature of difference schemes constructed by the above method is that grid points might not remain fixed in the course of the numerical integration. Precise criteria for a grid to be uniform, orthogonal and possessing flat time layers are formulated as conditions on the coefficients of infinitesimal symmetry generators and are broken, e.g., for Galilean boosts and inversions~\cite{doro11Ay}. This means that for such symmetries it is not possible to use isotropic or static grids. Hence, the problem of establishing good conditions governing the position of grid points both spatially and temporally becomes vital.

Up to now the reviewed technique has been applied to physically rather simple models, usually only involving time and one space dimension~\cite{baki97Ay,budd01Ay,doro03Ay,doro00Ay,vali05Ay}. It is understandable that the multidimensional case is even more delicate, as there is an increasing number of possibilities for assembling the difference invariants to finite difference schemes. In addition, grids can evolve differently in distinct spatial dimensions, which might cause severe numerical problems, such as tangling meshes, if not treated appropriately. In the present paper, we aim to discuss ways of overcoming the latter problem.

An alternative approach to constructing finite difference schemes with symmetry properties uses moving frames in the Fels and Olver formulation~\cite{fels98Ay}.
In contrast to the  Dorodnitsyn approach, where finite difference schemes are constructed from the outset, in the moving frame method the concept of \textit{invariantization} of existing schemes plays the key role. This technique can be summed up as follows~\cite{kim06Ay,kim08Ay}: Determine the Lie symmetry group of the given system of differential equations. This part is standard and usually involves exponentiation of elements of the maximal Lie invariance algebra of the system. Subsequently, a moving frame associated with the Lie symmetry group is constructed. Roughly speaking, a moving frame is an equivariant function that returns the unique group element mapping a given point to a point of a chosen submanifold (the cross-section), which intersects each group orbit once and transversally. Since the condition for submanifolds to be cross-sections to the group orbits is quite general, there is a freedom in choosing such submanifolds and hence in constructing the associated moving frames. Once a moving frame is obtained, it can be used to map an arbitrary function to an invariant function. This is a general property of any moving frame that allows determining usual invariants and differential invariants of a group action~\cite{olve03Ay,olve07Ay}. In the same way, it is possible to take a given difference scheme (considered as a function of grid points) for a system of differential equations and invariantize it by a moving frame. By this procedure, the given scheme is transformed to a new scheme that will be invariant under the same Lie group as was used to determine the moving frame.

The main benefit of this method is the possibility of using existing finite difference schemes as a starting point in the development of invariant schemes. Consequently, such invariant schemes could eventually be implemented in existing numerical models with limited effort.
At the same time, the freedom in constructing a moving frame can make it somewhat difficult to predict the precise form of invariantized expressions and, therefore, to arrive at a scheme that not only is invariant but also has some desirable numerical properties, as discussed, e.g., in~\cite{chha10Ay}.
Although assembling difference invariants to invariant discretization schemes can be realized in a variety of ways too, the form of the particular difference invariants usually imposes enough hints in order to find reasonable finite difference approximations of a given system of differential equations.
As with the Dorodnitsyn method described above, invariantization of existing numerical schemes may also lead to grids that evolve during numerical integration.

\looseness=-1
Another method for the construction of schemes with certain invariance properties was proposed in~\cite{budd96Ay} for equations describing blow-up problems; see also~\cite{budd09Ay,budd99Ay,huan10By}. The main idea in this approach is to use adaptive moving meshes from the very beginning because they are well suited for problems that develop shocks after a finite integration time. As a moving mesh complicates the discretization of differential equations in the physical space, the system to be discretized is first transformed into so-called \emph{computational coordinates} that remain orthogonal and do not evolve during the numerical integration. The physical system is then discretized in the computational coordinates. It is advocated in this approach that for equations exhibiting blow-ups and for the description of the solution near the singularity, scale invariance plays an exceptional role. Therefore, scale invariance is required to be preserved in the course of discretization. The evolution of the mesh is formulated as an auxiliary system of differential equations, the so-called \emph{moving mesh partial differential equations}. The auxiliary system is then selected in such a manner as to preserve the scale invariance of the original physical model. A straightforward extension to the above approach is to require the mesh equations to possess not only the scale symmetries but also the other symmetries that the system of physical differential equations admits. The following property is basic for this extension: \emph{The prolongation of any point symmetry of the initial system~$\mathcal L$ of differential equations to the computational coordinates by means of the identical transformation is a point symmetry of the counterpart of~$\mathcal L$ in terms of the computational coordinates.}

In the present paper, we will introduce yet another approach to the construction of invariant discretization schemes, which will be essential for multidimensional systems of differential equations. It rests on first expressing the system of differential equations under consideration in terms of computational coordinates and then extending the symmetry transformations of the original system to the system written in computational variables. Once it is understood how the system behaves under the extended symmetry transformations, one constructs a finite difference scheme that is transformed by the discretized version of the extended transformations in a similar way. In addition, the extra differential equations that control the location of the grid points are discretized in an invariant way, e.g., by using the finite difference invariants.

The reason why it is necessary to develop one more technique for the construction of invariant discretization schemes is twofold. Firstly, it is rather difficult to set up a proper invariant scheme for systems of differential equations using difference invariants as basic building blocks. For multidimensional moving meshes, an additional problem is to find proper finite difference analogs of derivatives. Secondly, it can be (and, in general, will be) desirable to include additional qualitative properties of differential equations in the construction of invariant discretization schemes. Within the invariantization technique it might be tedious to ensure the numerical preservation of certain conservation laws, even if the initial system includes equations represented as conservation laws, which is precisely the case for the shallow-water equations written in momentum form. A similar remark holds for the Dorodnitsyn approach. An exception is given for equations derived from a variational principle for which, in view of the discrete version of the Noether identity, conservation laws and symmetries can be simultaneously preserved in the course of a proper invariant discretization of the associated Lagrangian~\cite{budd01Ay}. There is still no algorithm using only difference invariants (or the invariantization map) that guarantees that the resulting invariant scheme will admit certain conservation laws. On the other hand, with the new approach to be introduced in the present paper, it is possible to construct, in a quite direct way, schemes that both are invariant and preserve some of the conservation laws possessed by the initial system.

\section{Invariant numerical models for the one-dimensional\texorpdfstring{\\}{ }shallow-water equations}\label{sec:OneDimensionalShallowWater}

In Section~\ref{sec:SymmetriesShallowWater} we discussed the Lie symmetries of the shallow-water equations without any relation to boundary value problems. However, when setting up a numerical model for a specific set of problems, the explicit treatment of certain boundary conditions is usually inevitable. As a rule, the Lie symmetries possessed by a boundary value problem form only a subgroup (often even trivial one) of the maximal Lie symmetry group admitted by the involved system of differential equations~\cite{blum89Ay}. Stated in another way, a specific boundary value problem usually admits only a small subset of the symmetries of the associated system of differential equations considered without boundary and initial conditions. This is in particular the case for various differential equations arising in hydrodynamics, which admit wide Lie invariance groups in the absence of boundary and initial conditions \cite{andr98Ay,bihl10Ay,bihl09Ay,bihl11By,ibra95Ay,ovsi82Ay}. As shown in the present section, this is also the case for the shallow-water equations. Therefore, it is necessary to find a way to incorporate the boundary and initial conditions considered into the numerical model to be developed.

\subsection{Selection of symmetries using boundary conditions}

In order to design invariant numerical schemes for a system of differential equations, two principal strategies can be adopted.

In the first approach, which is applied in most of the previous works on invariant discretization~\cite{doro11Ay}, numerical schemes preserving the entire maximal Lie invariance groups of the corresponding systems are developed and then implemented for the specific physical configurations of interest. The drawback of this approach is that the practical implementation of a numerical scheme always requires the explicit treatment of a boundary value problem. As was said above, for the boundary conditions arising most often in hydrodynamics (e.g., periodic, reflective or absorbing), the maximal Lie invariance groups of systems without boundary conditions are usually much wider than those of particular boundary value problems. Using the first approach may therefore lead to the overly restrictive requirement that all the symmetries of the considered system of differential equations are equally important in the course of invariant discretization. For a particular boundary value problem, however, this may not be the case, as some of the symmetries of the system might not be admitted at all.

This is why we adopt the second approach here, which only requires the preservation of symmetries that are compatible with the class of specific boundary value problems under consideration. The apparent drawback of this approach is that, if one aims to test different kinds of boundary conditions, it can be necessary to design a new scheme for each configuration, as different symmetry groups may arise when varying specific settings. On the other hand, for a model to be used for a particular purpose (e.g., a weather or climate prediction model), the boundary conditions are generally fixed at the stage of model development and therefore do not change subsequently. Another advantage of the second approach is of more physical nature. The restriction imposed on symmetries in that they map a given boundary value problem to itself is unreasonable even from the physical perspective. When transforming a given reference frame to another reference frame, the boundary value problems of the reference frames involved are also mapped to each other. Therefore, it is not natural to require appropriate symmetries of a system of differential equations to preserve a particular boundary value problem but rather only to impose that these symmetries map boundary value problems from a class of such problems (e.g., periodic domains of \emph{any} size with varying initial time and initial conditions) to each other. Such transformations are known as \emph{equivalence transformations}, and when deriving symmetry-preserving discretization schemes, we require a subgroup of the maximal Lie invariance group of the original system of differential equations to be compatible with the structure of a predefined class of boundary value problems, i.e., elements of the subgroup should act as equivalence transformations on the boundary conditions rather than as symmetry transformations.

The relaxed condition of requiring the finite difference schemes to be invariant only under the transformations admitted by a class of boundary value problems thus provides a natural selection criterion for subgroups of a maximal Lie invariance group to be preserved numerically. In view of the particular nature of the infinite-dimensional maximal Lie invariance algebra~$\mathfrak g_1$ of the one-dimensional shallow-water equations (or, more generally, systems of differential equations arising in hydrodynamics), it might be a cumbersome or even useless task to attempt to preserve all these symmetries in the respective discrete models.

Among the most natural boundary conditions for the one-dimensional shallow-water equations in the setting of geophysical fluid dynamics are periodic ones, which we aim to study here. These  boundary conditions are advantageous as they are not as restrictive as, e.g., Dirichlet boundary conditions from the pure symmetry point of view and generally lead to the selection of symmetries that have a clear physical interpretation.
The subalgebra~$\mathfrak s_1$ of~$\mathfrak g_1$ that is compatible with periodic boundary conditions is spanned by the vector fields
\begin{equation}\label{eq:SymmetriesBoundaryValueProblemOneDimensionalShallowWater}
    \p_t, \quad \p_x,\quad t\p_x+\p_u, \quad t\p_t+x\p_x,\quad x\p_x+u\p_u+2h\p_h.
\end{equation}
Even if we neglected initial conditions, only elements from the narrower subalgebra $\langle\p_t, \p_x,$ \mbox{$t\p_x+\p_u,$} $t\p_t-u\p_u-2h\p_h\rangle$ generate one-parameter symmetry groups of such a boundary value problem, as scalings with respect to~$x$ are not admitted once a domain (periodicity) length is fixed. However, the inclusion of scaling symmetries in the subsequent consideration is justified as they are equivalence transformations of the chosen class of boundary value problems. In other words, upon preserving the subalgebra~$\mathfrak s_1$ we are still able to test different domain lengths. In this sense, the preservation of scaling transformations plays an important role for the class of boundary value problems for the one-dimensional shallow-water equations with periodic boundary conditions and \textit{any} domain size even if scalings are not proper symmetry transformations for a specific numerical integration.

\subsection{Classical invariant schemes and beyond}\label{sec:ClassicalInvariantNumericalSchemes}

\begin{figure}[ht!]
\centering
\includegraphics{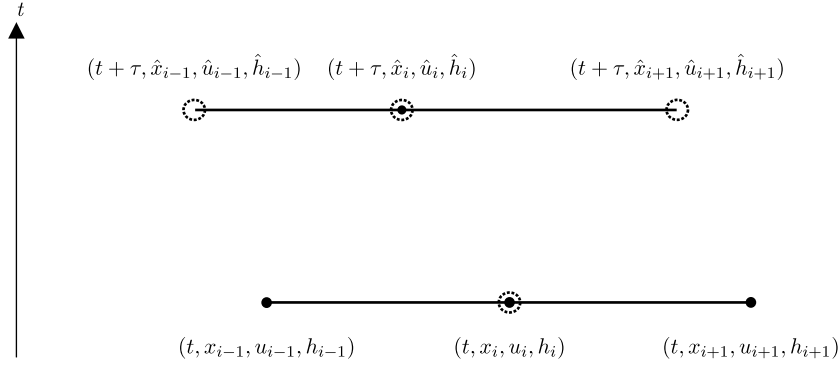}
\caption{Stencils for invariant schemes of the one-dimensional shallow-water equations. An explicit (Euler forward) and an implicit (Euler backward) scheme is defined using the points indicated by filled and dashed circles, respectively.}
\label{fig:StencilOneDim}
\end{figure}

We begin our study of invariant numerical schemes for the shallow-water equations using the classical construction proposed by Dorodnitsyn. Within this framework, we have to prolong the selected subalgebra~$\mathfrak s_1$ to the discretization stencils that we aim to use. These stencils are depicted in Fig.~\ref{fig:StencilOneDim}. As the symmetry group associated with~$\mathfrak s_1$ does not violate the criterion for using flat time layers (see~\cite{doro11Ay} for more details), all points in the spatial domain are defined at the same time. However, if we wish to preserve Galilean invariance in a numerical scheme, it is impossible to use a fixed grid, i.e., $\hat x_i\ne x_i$.
Here and in what follows variables with a hat and without a hat denote values on the grid at the time levels $t+\tau$ and~$t$, respectively, and $\tau$ is the time step.
The possibility of evolving grids in general also leads to nonhomogeneous spacings in the course of the integration; i.e., $\hat x_{i+1}-\hat x_{i}\ne \hat x_{i} - \hat x_{i-1}$ even if the initial grid $\{x_i\}$ is equally spaced.

Prolonging the vector fields~\eqref{eq:SymmetriesBoundaryValueProblemOneDimensionalShallowWater} to the points indicated in Fig.~\ref{fig:StencilOneDim} gives
\begin{align}\label{eq:ProlongedOperatorsOneDimensional}
\begin{split}
  &\p_t, \quad \p_{x_i}+\p_{x_{i+1}}+\p_{x_{i-1}}+\p_{\hat x_i}+\p_{\hat x_{i+1}}+\p_{\hat x_{i-1}},\\[1ex]
  &t(\p_{x_i}+\p_{x_{i+1}}+\p_{x_{i-1}})+(t+\tau)(\p_{\hat x_i}+\p_{\hat x_{i+1}}+\p_{\hat x_{i-1}})+{}\\
  &{}\p_{u_i}+\p_{u_{i+1}}+\p_{u_{i-1}}+\p_{\hat u_i}+\p_{\hat u_{i+1}}+\p_{\hat u_{i-1}}, \\[1ex]
  &t\p_t+\tau\p_\tau+x_i\p_{x_i}+x_{i+1}\p_{x_{i+1}}+x_{i-1}\p_{x_{i-1}}+\hat x_i\p_{\hat x_i}+\hat x_{i+1}\p_{\hat x_{i+1}}+\hat x_{i-1}\p_{\hat x_{i-1}},\\[1ex]
  & x_i\p_{x_i}+x_{i+1}\p_{x_{i+1}}+x_{i-1}\p_{x_{i-1}}+\hat x_i\p_{\hat x_i}+\hat x_{i+1}\p_{\hat x_{i+1}}+\hat x_{i-1}\p_{\hat x_{i-1}}+{}\\
  &{} u_i\p_{u_i}+u_{i+1}\p_{u_{i+1}}+u_{i-1}\p_{u_{i-1}}+\hat u_i\p_{\hat u_i}+\hat u_{i+1}\p_{\hat u_{i+1}}+\hat u_{i-1}\p_{\hat u_{i-1}}+{}\\
  &{} 2h_i\p_{h_i}+2h_{i+1}\p_{h_{i+1}}+ 2h_{i-1}\p_{h_{i-1}}+2\hat h_i\p_{\hat h_i}+ 2\hat h_{i+1}\p_{\hat h_{i+1}}+2\hat h_{i-1}\p_{\hat h_{i-1}}.
\end{split}
\end{align}

To construct an explicit (Euler forward) numerical scheme, we have to restrict ourselves in~\eqref{eq:ProlongedOperatorsOneDimensional} to the
values that are defined at the filled circles depicted in Fig.~\ref{fig:StencilOneDim}. A~convenient complete set of functionally independent difference invariants is then
\begin{align}\label{eq:DifferenceInvariantsExplicitSchemeOneDimensional}
\begin{split}
 &I_0=\frac{x_{i+1}- x_{i}}{x_{i}-x_{i-1}},\quad I_1=\frac{\dot x_i-u_i}{x_{i+1}-x_{i-1}}\tau, \quad I_2=\frac{\hat u_i - u_i}{x_{i+1}-x_{i-1}}\tau,\\[.3ex]
 &I_3=\frac{u_{i+1}-u_{i-1}}{x_{i+1}-x_{i-1}}\tau,\quad I_4=\frac{u_{i+1}-u_{i}}{x_{i+1}-x_{i}}\tau,\quad I_5=\frac{h_{i-1}}{(x_{i+1}-x_{i-1})^2}\tau^2,\\[.3ex]
 &I_6 = \frac{h_i}{(x_{i+1}-x_{i-1})^2}\tau^2,\quad I_7=\frac{h_{i+1}}{(x_{i+1}-x_{i-1})^2}\tau^2,\quad  I_8=\frac{\hat h_i}{(x_{i+1}-x_{i-1})^2}\tau^2,
\end{split}
\end{align}
where $\dot x_i= (\hat x_i - x_i)/\tau$ is by definition the mesh velocity.
These invariants are found from integrating the system of first-order quasilinear partial differential equations $\vv_j (I)=0$, where $\vv_j$, $j=1,\dots,5$, are the prolonged vector fields presented in~\eqref{eq:ProlongedOperatorsOneDimensional}. The determining system for invariants admits precisely $m_{\rm i}=m_{\rm s}-r$ functionally independent solutions, where $m_{\rm s}$ is the number of stencil variables and $r$ is the rank of involved vector fields. For the explicit scheme, we have $m_{\rm s}=14$, $r=5$ and hence $m_{\rm i}=9$.

Using the difference invariants of the set~\eqref{eq:DifferenceInvariantsExplicitSchemeOneDimensional}, we can approximate~\eqref{eq:OneDimensionalShallowWater} via
\begin{align*}
 &I_1=0,\quad I_2+I_7-I_5=0,\quad I_8-I_6+I_6I_3=0
\end{align*}
or, explicitly,
\begin{align}\label{eq:ExplicitInvariantSchemeOneDimensionalShallowWater}
\begin{split}
 &\dot x_i = u_i, \quad
\frac{\hat u_i-u_i}{\tau} + \frac{h_{i+1}-h_{i-1}}{x_{i+1}-x_{i-1}}=0, \quad
\frac{\hat h_i-h_i}{\tau} + h_i\frac{u_{i+1}-u_{i-1}}{x_{i+1}-x_{i-1}}=0.
\end{split}
\end{align}
In the continuous limit the above scheme leads to the following system of differential equations:
\begin{equation}\label{eq:ContinuousLimitOneDimensional}
 \dd{x}{t}=u,\quad \dd{u}{t}+\pdl{h}{x}=0,\quad \dd{h}{t}+h\pdl{u}{x}=0,
\end{equation}
which is~\eqref{eq:OneDimensionalShallowWater} in Lagrangian variables.

As the Euler forward scheme is only conditionally stable, it is beneficial to construct an implicit invariant numerical scheme. A simple implicit scheme is the Euler backward scheme, which can be constructed in a similar way as the invariant Euler forward scheme. However, we prefer to at once construct a trapezoidal scheme, which has in general a greater accuracy. To accomplish this we additionally need the difference invariants
\begin{align}\label{eq:DifferenceInvariantsImplicitSchemeOneDimensional}
\begin{split}
 &I_9=\frac{\dot x_i-\hat u_i}{x_{i+1}-x_{i-1}}\tau, \quad I_{10}=\frac{\hat u_{i+1}-\hat u_{i-1}}{\hat x_{i+1}-\hat x_{i-1}}\tau,\quad
 I_{11} = \frac{\hat h_{i+1}-\hat h_{i-1}}{(x_{i+1}-x_{i-1})(\hat x_{i+1}-\hat x_{i-1})}\tau^2.
\end{split}
\end{align}
Note that $\{I_0,\dots,I_{11}\}$ is not a complete set of functionally independent invariants for the transformation group generated by the vector fields~\eqref{eq:ProlongedOperatorsOneDimensional} on the trapezoidal stencil as the total number of variables on this stencil equals 20 and hence a functional basis of related invariants consists of 15 invariants.
By combining the invariants $I_1$--$I_3$ and $I_5$--$I_{11}$ we construct
\begin{align*}
 & I_1+I_9=0,\quad I_2+\frac12(I_7-I_5+I_{11})=0,\quad I_8-I_6 +\frac12(I_6I_3+I_8I_{10})=0,
\end{align*}
which boils down to the form
\begin{align}\label{eq:ImplicitInvariantSchemeOneDimensionalShallowWater}
\begin{split}
 &\dot x_i = \frac12(u_i+\hat u_i), \\
 &\frac{\hat u_i-u_i}{\tau} + \frac12\left(\frac{h_{i+1}-h_{i-1}}{x_{i+1}-x_{i-1}}+\frac{\hat h_{i+1}-\hat h_{i-1}}{\hat x_{i+1}-\hat x_{i-1}}\right)=0, \\
 &\frac{\hat h_i-h_i}{\tau} + \frac12\left(h_i\frac{u_{i+1}-u_{i-1}}{x_{i+1}-x_{i-1}}+ \hat h_i\frac{\hat u_{i+1}-\hat u_{i-1}}{\hat x_{i+1}-\hat x_{i-1}}\right)=0.
\end{split}
\end{align}
This scheme also converges to~\eqref{eq:ContinuousLimitOneDimensional}.

The problem with the above schemes in particular and with the difference invariants approach to the construction of invariant schemes in general is that it is hard to control properties other than symmetries that the resulting discretizations admit. It can be checked by direct computation that the above schemes violate even the mass conservation law (conservation of momentum and energy is violated as well). This violation of fundamental conservation laws is a direct consequence of the construction method of the invariant finite difference schemes, which only takes into account local information on $u_i$ and $h_i$ (i.e., the difference invariants) but provides no guideline ensuring the preservation of global features by the numerical solution.

In the present case, this problem can be partially circumvented by discretizing the shallow-water equations not in the form~\eqref{eq:ContinuousLimitOneDimensional} but rather in the momentum form
\begin{equation}\label{eq:ShallowWaterConservedFormOneDimensional}
 \mathrm{Eq}^h=h_t+(uh)_x=0,\quad
 \mathrm{Eq}^u=(uh)_t +\left(hu^2+\frac12h^2\right)_x=0. 
\end{equation}
An invariant finite difference approximation of this system using an Euler time step is
\begin{align*}
 &\dot x_i = u_i,\quad 
 \hat h_i\dfrac{\hat x_{i+1}-\hat x_{i-1}}{x_{i+1}-x_{i-1}}-h_i=0,\quad
 \hat u_i\hat h_i\dfrac{\hat x_{i+1}-\hat x_{i-1}}{x_{i+1}-x_{i-1}}-u_ih_i +\frac\tau2\frac{h_{i+1}^2-h_{i-1}^2}{x_{i+1}-x_{i-1}}=0, 
\end{align*}
while the invariant trapezoidal scheme is
\begin{align}\label{eq:ImplicitInvariantSchemeOneDimensionalShallowWaterConservative}
\begin{split}
 &\dot x_i = \frac12(u_i+\hat u_i),\quad 
 \hat h_i\dfrac{\hat x_{i+1}-\hat x_{i-1}}{x_{i+1}-x_{i-1}}-h_i=0,\\[1ex]
 &\hat u_i\hat h_i\dfrac{\hat x_{i+1}-\hat x_{i-1}}{x_{i+1}-x_{i-1}}-u_ih_i +\frac\tau4\left(\frac{h_{i+1}^2-h_{i-1}^2}{x_{i+1}-x_{i-1}}+\frac{\hat h_{i+1}^2-\hat h_{i-1}^2}{\hat x_{i+1}-\hat x_{i-1}}\right)=0.
\end{split}
\end{align}
The schemes constructed in this way numerically preserve the mass conservation law. The explicit scheme in addition preserves the momentum, while the implicit scheme~\eqref{eq:ImplicitInvariantSchemeOneDimensionalShallowWaterConservative} does not preserve momentum exactly, as $\hat x_{i+1}-\hat x_{i-1}\ne x_{i+1}- x_{i-1}$ in general. This condition, however, would be required to yield exact conservation of momentum but, as the change in the spacings $x_{i+1}- x_{i-1}$ is not abrupt, the violation of momentum conservation of the scheme~\eqref{eq:ImplicitInvariantSchemeOneDimensionalShallowWaterConservative} is rather small; see the result of a numerical integration using~\eqref{eq:ImplicitInvariantSchemeOneDimensionalShallowWaterConservative} below. None of these schemes respects conservation of energy.

In the continuous limits both the explicit and implicit discretizations give
\[
 \dd{x}{t}=u,\quad \dd{h}{t}+h\pdl{u}{x}=0,\quad \dd{(uh)}{t}+\frac12\pdl{h^2}{x}=0,
\]
which is~\eqref{eq:ShallowWaterConservedFormOneDimensional} expressed in Lagrangian variables.
Similar as the schemes~\eqref{eq:ExplicitInvariantSchemeOneDimensionalShallowWater} and~\eqref{eq:ImplicitInvariantSchemeOneDimensionalShallowWater} the above two invariant schemes for the momentum form of the shallow-water equations could be expressed in terms of difference invariants. As these expressions are considerably more involved than the analogous expressions for~\eqref{eq:ExplicitInvariantSchemeOneDimensionalShallowWater} and~\eqref{eq:ImplicitInvariantSchemeOneDimensionalShallowWater}, we do not present these difference invariant forms here. The reason for the difference invariant expressions being more complicated in the present case can be traced back to the result of the Galilean transformation $\tilde t=t$, $\tilde x=x+\ve_1t$, $\tilde u=u+\ve_1$, when applied to~\eqref{eq:ShallowWaterConservedFormOneDimensional}, which yields
\begin{equation}\label{eq:GalileanTransformation1d}
\widetilde{\rm Eq}{}^h={\rm Eq}^h,\quad
\widetilde{\rm Eq}{}^u={\rm Eq}^u+\varepsilon_1{\rm Eq}^h.
\end{equation}
The transformation of the momentum equation thus yields a linear combination of the momentum equation with the continuity equation. This implies that the momentum equation itself cannot be expressed in terms of differential invariants but only in combination with the continuity equation. It is thus not natural to approximate the momentum equation using difference invariants. At the same time, checking that the proposed conservative schemes are indeed invariant can be shown directly by acting with the prolonged vector fields~\eqref{eq:ProlongedOperatorsOneDimensional} on them and verifying that the results of these operations yield zero on the solutions of the numerical scheme.

\begin{figure}[ht!]
\centering
\includegraphics[scale=1.05]{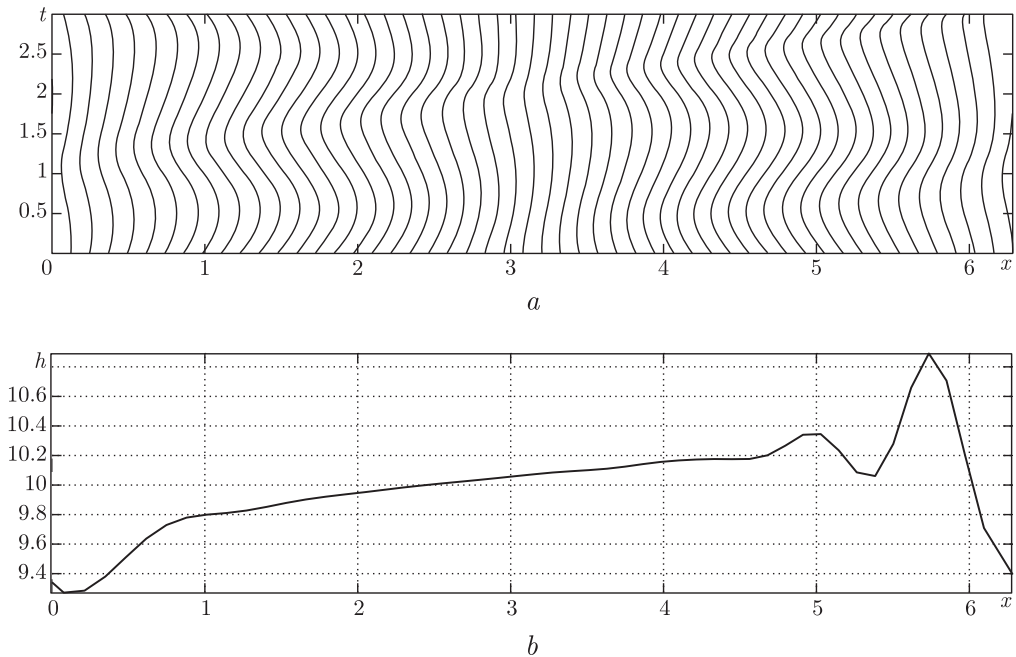}
\caption{Numerical integration of the one-dimensional shallow-water equations~\eqref{eq:OneDimensionalShallowWater} using the scheme~\eqref{eq:ImplicitInvariantSchemeOneDimensionalShallowWater} with $\tau=0.001$ and $N=51$ grid points on the domain $[0, 2\pi]$ over the time interval $[0,3]$. The initial conditions are $u=A\sin x$ and $h=h_0+A\sin(x+\varphi_0)$, with $A=0.4$, $\varphi_0=\pi/6$ and $h_0=10$.
({\it a}) Evolution of the discretization grid.
({\it b}) Numerical solution for $h$ at $t=3$.}
\label{fig:LagrangianImplicitInvariantSchemeOneDimensionalShallowWaterResult}
\end{figure}

The result of a numerical integration taking harmonic initial conditions for both~$u$ and~$h$ using the scheme~\eqref{eq:ImplicitInvariantSchemeOneDimensionalShallowWaterConservative} is depicted in Fig.~\ref{fig:LagrangianImplicitInvariantSchemeOneDimensionalShallowWaterResult}. As the evolution of the mesh points is directly coupled to the (initially harmonic) physical velocity, the single mesh points quasi-oscillate around their initial positions (Fig.~\ref{fig:LagrangianImplicitInvariantSchemeOneDimensionalShallowWaterResult}{\it a}). No special ability of the mesh to follow the developing shock (Fig.~\ref{fig:LagrangianImplicitInvariantSchemeOneDimensionalShallowWaterResult}{\it b}, showing the numerical solution of~$h$ at time $t=3$) is visible, which is one of the major disadvantages of the scheme~\eqref{eq:ImplicitInvariantSchemeOneDimensionalShallowWaterConservative}. The scheme conserves mass up to machine precision ($10^{-16}$) but it dissipates energy, with the relative change $(\mathcal H(t)-\mathcal H(0))/\mathcal H(0)$ of energy being of the order $10^{-5}$ at the end of the integration. The relative change in momentum in this integration is of order $10^{-14}$ without a positive or negative trend. The values of~$\mathcal M$, $\mathcal H$ and~$\mathcal P$ at time~$t$ are evaluated using the formulas
$\mathcal M=\frac12\sum_ih_i(x_{i+1}-x_{i-1})$, $\mathcal H=\frac14\sum_i(h_iu_i^2+h_i^2)(x_{i+1}-x_{i-1})$
and~$\mathcal P_=\frac12\sum_ih_iu_i(x_{i+1}-x_{i-1})$, respectively.

\subsection{Invariant discretization on equidistributing meshes}\label{sec:AlternativeSolutionProcedure}

\looseness=1
So far, we have mainly been concerned with assembling difference invariants in a proper way, so as to guarantee the invariance of the resulting finite difference schemes. That is, the invariance condition was the relevant starting point in the design of the above schemes. The main problem with this approach is the lack of an explicit error control for the proposed numerical models. When setting up a numerical scheme, as well as being of primary interest to ensure the discrete preservation of qualitative features of differential equations, it is also important to address classical issues from numerical analysis. For adaptive moving meshes, these issues mainly concern the prevention of abruptly changing grids, mesh racing and mesh tangling, which can significantly degrade the numerical solution and ultimately lead to convergence failure.

When dealing with finite difference schemes on adaptive moving meshes, one usually regards the mesh movement as a time-dependent coordinate transformation from a fixed logical (computational) domain to the physical domain of the system of differential equations. The computational coordinates are defined to index the positions of the grid points in the mesh. Because in any regular grid each grid point keeps its position relative to its neighbors in the mesh even in the presence of adaptation, it is convenient to take the (spatial) computational coordinates as time-independent, Cartesian and orthogonal, with uniform spacing on the unit interval (up to scaling). In the one-dimensional case considered here, the step of the spatial computational coordinate $\xi$ equals $1/(N-1)$, where $N$ is the number of grid points at each fixed time level. In order to use computational coordinates, it is necessary to transform the system of differential equations from the physical space to the index space; see, e.g.,~\cite{budd09Ay,huan10By}.

The relation of the usage of computational coordinates to invariant numerical schemes will be illustrated again with the one-dimensional system of shallow-water equations. The central idea is that any finite difference discretization of the shallow-water equations on a moving mesh in computational coordinates is invariant under the Lie group generated by the vector fields~\eqref{eq:ProlongedOperatorsOneDimensional}. Indeed, under the transformation $t=\theta$, $x=x(\theta,\xi)$ to the computational coordinates $(\theta,\xi)$, the one-dimensional shallow-water equations~\eqref{eq:OneDimensionalShallowWater} take the form
\begin{align}\label{eq:1DShallowWaterComputationalCoordinates}
&\tilde u_\theta + (\tilde u-x_\theta)\frac{\tilde u_\xi}{x_\xi} + \frac{1}{x_\xi}\tilde h_\xi=0,\quad
 \tilde h_\theta + (\tilde u-x_\theta)\frac{\tilde h_\xi}{x_\xi} + \frac{1}{x_\xi}\tilde h\tilde u_\xi=0,
\end{align}
where $\tilde u=u(\theta,x(\theta,\xi))$ and $\tilde h=h(\theta,x(\theta,\xi))$.
It is obvious that any usual finite difference discretization of~\eqref{eq:1DShallowWaterComputationalCoordinates} possesses the symmetry group requested.
For example, discretizing using forward differences in time and central differences in space, from the above system we obtain
\begin{align*}
\begin{split}
 &\frac{\hat u_i-u_i}{\tau} +\left(u_i-\dot x_i\right)\frac{u_{i+1}-u_{i-1}}{x_{i+1}-x_{i-1}} + \frac{h_{i+1}-h_{i-1}}{x_{i+1}-x_{i-1}}=0,
 \\
 &\frac{\hat h_i-h_i}{\tau} +\left(u_i-\dot x_i\right)\frac{h_{i+1}-h_{i-1}}{x_{i+1}-x_{i-1}} + h_i\frac{u_{i+1}-u_{i-1}}{x_{i+1}-x_{i-1}}=0,
\end{split}
\end{align*}
where $u_i = \tilde u(\theta,\xi_i)=u(t,x_i(t))$, $h_i = \tilde h(\theta,\xi_i)=h(t,x_i(t))$ and $\hat u_i$ and~$\hat h_i$ denote the same values at $\theta+\tau$.
This discretization coincides with the second and third equations of the system~\eqref{eq:ExplicitInvariantSchemeOneDimensionalShallowWater} if we assume the grid evolution to be Lagrangian of the form $\dot x_i = u_i$.

In much the same way, an invariant implicit discretization (trapezoidal rule) can be obtained from~\eqref{eq:1DShallowWaterComputationalCoordinates}, giving
\begin{align*}
\begin{split}
 &\frac{\hat u_i-u_i}{\tau} +\frac12\left(\frac{u_i+\hat u_i}2-\dot x_i\right)\left(\frac{u_{i+1}-u_{i-1}}{x_{i+1}-x_{i-1}}+ \frac{\hat u_{i+1}-\hat u_{i-1}}{\hat x_{i+1}-\hat x_{i-1}}\right) + {}\\ &{}\frac12\left(\frac{h_{i+1}-h_{i-1}}{x_{i+1}-x_{i-1}}+\frac{\hat h_{i+1}-\hat h_{i-1}}{\hat x_{i+1}-\hat x_{i-1}}\right)=0,
 \\
 &\frac{\hat h_i-h_i}{\tau} +\frac12\left(\frac{u_i+\hat u_i}2-\dot x_i\right)\left(\frac{h_{i+1}-h_{i-1}}{x_{i+1}-x_{i-1}} + \frac{\hat h_{i+1}-\hat h_{i-1}}{\hat x_{i+1}-\hat x_{i-1}}\right) + {}\\ &{}\frac12\left(h_i\frac{u_{i+1}-u_{i-1}}{x_{i+1}-x_{i-1}}+ \hat h_i\frac{\hat u_{i+1}-\hat u_{i-1}}{\hat x_{i+1}-\hat x_{i-1}}\right)=0.
\end{split}
\end{align*}
Again, in the Lagrangian case $\dot x_i = (u_i+\hat u_i)/2$, this scheme coincides with the scheme~\eqref{eq:ImplicitInvariantSchemeOneDimensionalShallowWater}.

Neither of these two schemes preserves mass and momentum, as they approximate the representation~\eqref{eq:1DShallowWaterComputationalCoordinates} of the shallow-water equations, where the equations are not in conserved form. It is possible to discretize the conserved form~\eqref{eq:ShallowWaterConservedFormOneDimensional} using computational variables as well, which boils down to
\begin{align*}
 &\hat h_i\dfrac{\hat x_{i+1}-\hat x_{i-1}}{x_{i+1}-x_{i-1}}-h_i-\tau A(h)=0,\quad
 \hat u_i\hat h_i\dfrac{\hat x_{i+1}-\hat x_{i-1}}{x_{i+1}-x_{i-1}}-u_ih_i +\tau A(uh)+\frac\tau2D(h^2)=0
\end{align*}
for the explicit Euler scheme (preserving mass and momentum) and to
\begin{align}\label{eq:InvariantSchemeShallowWaterComputationalCoordinates}
\begin{split}
 &\hat h_i\dfrac{\hat x_{i+1}-\hat x_{i-1}}{x_{i+1}-x_{i-1}}-h_i-\frac\tau2(A(h)+\hat A(h))=0,
\\
 &\hat u_i\hat h_i\dfrac{\hat x_{i+1}-\hat x_{i-1}}{x_{i+1}-x_{i-1}}-u_ih_i +\frac\tau2 (A(uh)+\hat A(uh))+\frac\tau4(D(h^2)+\hat D(h^2))=0 
\end{split}
\end{align}
for the implicit trapezoidal discretization. In both schemes we denote
\[
  A(z)=\frac{(u_{i+1}-\dot x_{i+1})z_{i+1}-(u_{i-1}-\dot x_{i-1})z_{i-1}}{x_{i+1}-x_{i-1}},\quad D(z)= \frac{z_{i+1}-z_{i-1}}{x_{i+1}-x_{i-1}},
\]
and $\hat A(z)$ and $\hat D(z)$ have the same forms as $A(z)$ and $D(z)$ with all the variables replaced by the associated variables on the next time step $\theta +\tau$, only keeping the grid velocity the same. In the continuous limit, these schemes converge to
\[
 (x_\xi F^t)_\theta+(F^x-F^tx_\theta)_\xi=0,
\]
which is indeed~\eqref{eq:ShallowWaterConservedFormOneDimensional} in computational coordinates using $F^t=(h,hu)$ and $F^x=(hu,hu^2+\tfrac12h^2)$.

The above observation can be easily extended to other invariant schemes for evolution equations admitting Galilean transformations as symmetries. Its main benefit is that it allows us to establish a connection to the theory of discretization on adaptive moving meshes. This may aid in tackling the problem of finding invariant finite difference schemes that also have good numerical properties.

In order to complete the invariant schemes in computational coordinates, it is necessary to determine the mesh velocity $\dot x_i$ in an invariant way. This can be done using \textit{equidistributing meshes}. Classically, a mesh is called equidistributed if the relation
\[
 \int_{a}^{x(\xi)}\rho(x)\ddd x = \xi\int_{a}^{b}\rho(x)\ddd x
\]
holds for the continuous mapping $x=x(\xi)\colon [0, 1]\to[a,b]$; see, e.g.,~\cite{huan10By}. The function $\rho=\rho(x)$ is called the monitor function. It determines the regions of concentration of the grid. Differentiating this relation twice with respect to $\xi$, one obtains the equation
\begin{equation}\label{eq:EquidistributionPrinciple}
 (\rho x_\xi)_\xi = 0,
\end{equation}
with the boundary conditions $x(0)=a$ and $x(1)=b$, which is satisfied for an equidistributed mesh. As we are considering periodic boundary conditions, we should modify the classical framework of equidistributing meshes and replace the boundary conditions for $x(\xi)$ by setting $x(1)-x(0)=2\pi$ and $x_\xi(0)=x_\xi(1)$. The periodic conditions for $x(\xi)$ agree with the invariance requested.

The above schemes in computational coordinates will therefore be completely invariant if we obtain the grid on the next level (and therefore the grid velocity $\dot x_i= (\hat x_i-x_i)/\tau$) from an invariant discretization of the equidistribution principle~\eqref{eq:EquidistributionPrinciple}. The discretization
\begin{equation}\label{eq:DiscretizationEquidistributionPrinciple}
 (\rho_{i+1}+\rho_i)(\hat x_{i+i}-\hat x_{i})-(\rho_i+\rho_{i-1})(\hat x_i-\hat x_{i-1})=0
\end{equation}
is invariant provided that we choose an invariant monitor function~$\rho$. An ansatz for~$\rho$ motivated from the theory of adaptive grids is, e.g., the arc-length(-like) monitor function
\[
 \rho = \sqrt{1+\alpha u_x^2}
\]
with $\alpha$ being the (positive) adaptation constant.
This monitor function is invariant with respect to vector fields~\eqref{eq:SymmetriesBoundaryValueProblemOneDimensionalShallowWater}
excluding only the scale operator $t\p_t+x\p_x$ but the corresponding scalings are equivalence transformations for the set of such monitor functions, where the parameter~$\alpha$ varies.
The above ansatz for~$\rho$ can be discretized in an invariant way via
\begin{equation}\label{eq:DiscretizationInvariantMonitorFunction}
 \rho_i = \sqrt{1+\alpha \left(\frac{u_{i+1}- u_{i-1}}{x_{i+1}-x_{i-1}}\right)^2}.
\end{equation}
The resulting form of~\eqref{eq:DiscretizationEquidistributionPrinciple} can then be solved either using an iterative method, such as Jacobi or Gau\ss--Seidel iteration, or by relaxation, e.g., using the moving mesh PDE approach~\mbox{\cite{budd09Ay,huan10By}}.

\begin{remark}\label{re:OnChoice ofInvMonitorFunctions1D}\looseness=-1
For the equation~\eqref{eq:EquidistributionPrinciple} to possess a Lie symmetry algebra~$\mathfrak g$ that is contained in the linear span~$\mathfrak s_1$ of vector fields~\eqref{eq:SymmetriesBoundaryValueProblemOneDimensionalShallowWater} trivially extended to~$\xi$, it suffices for the monitor function~$\rho$ to be an invariant of~$\mathfrak g$. On solutions of the shallow-water equations~\eqref{eq:OneDimensionalShallowWater} we can assume without loss of generality that the function~$\rho$ does not depend on derivatives of~$u$ and~$h$ involving differentiation with respect to~$t$. Then the general form of~$\rho$ that is an invariant of the pure Galilean algebra $\langle\p_t,\p_x,t\p_x+\p_u\rangle$ is given by an arbitrary smooth function of derivatives of~$u$ and~$h$ with respect to~$x$ including $h$ itself but not~$u$. In order to attain invariance with respect to scale transformations, the function~$\rho$ should depend only on specific products of powers of the above derivatives. At the same time, the incorporation of geometric properties of solutions (e.g., the length of a graph between neighboring grid points) into the monitor function is more important than scale invariance. Therefore, scale transformations can be allowed to act in a relaxed way, as equivalence transformations on a selected narrowed set of monitor functions. An obvious form that satisfies this requirement is the arc-length monitor function $\rho = \sqrt{1+\alpha u_x^2}$. Alternatively, one could use, e.g., the similar functions $\rho=\sqrt{1+\alpha h_x^2}$ and $\rho=\sqrt{1+\alpha u_x^2+\beta h_x^2}$ or the curvature-related monitor functions $\rho=\sqrt{1+\alpha u_{xx}^2}$, $\rho=\sqrt{1+\alpha h_{xx}^2}$ and $\rho=\sqrt{1+\alpha u_{xx}^2+\beta h_{xx}^2}$, where $\alpha$ and $\beta$ are positive constants.
\end{remark}

\begin{remark}
The method of constructing an invariant discretization of a differential equation in combination with a numerical grid generator was discussed, e.g., in \cite{budd98By,budd98Ay,budd99Ay}. In contrast to the method employed above, in~\cite{chha11Ay} the space of stencil variables was also prolonged to the monitor function, which is not necessarily chosen in an invariant way. In order to arrive at a completely invariant model, we however regard it important that the equidistribution principle is discretized in an invariant fashion too; see also the discussion in Section~\ref{sec:Conclusion}. Moreover, as the monitor function involves independent variables, unknown functions and their derivatives, it is possible to express its discretization using the same basis difference invariants of stencil variables that is needed for the physical differential equation discretization. In other words, no explicit prolongation to the monitor function is necessary within the framework of our approach.
\end{remark}

\begin{figure}[ht!]
\centering
\includegraphics{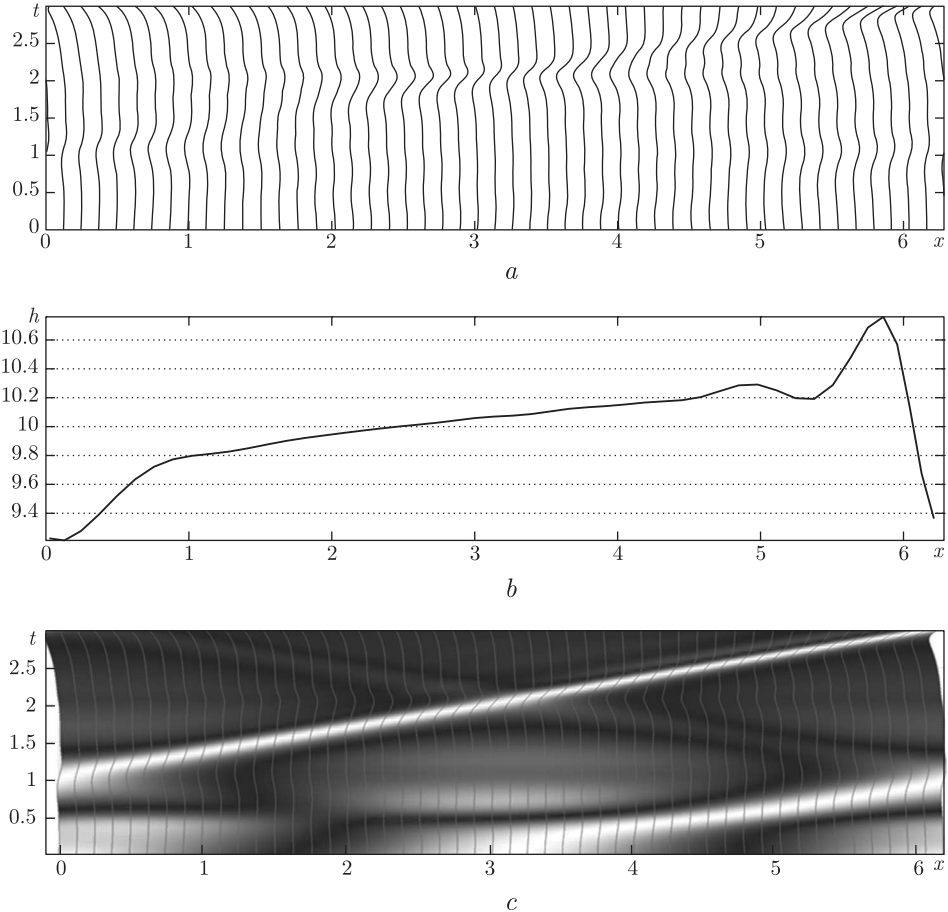}
\caption{Numerical integration of the one-dimensional shallow-water equations~\eqref{eq:OneDimensionalShallowWater} using the scheme~\eqref{eq:InvariantSchemeShallowWaterComputationalCoordinates} with $\tau=0.001$ and $N=51$ grid points on the domain $[0, 2\pi]$ over the time interval $[0,3]$. The initial conditions are $u=A\sin x$ and $h=h_0+A\sin(x+\varphi_0)$, with $A=0.4$, $\varphi_0=\pi/6$ and $h_0=10$. The trapezoidal rule is used for time integration and the arc-length monitor function is chosen for grid adaptation, setting $\alpha=0.8$.
({\it a}) Evolution of the discretization grid.
({\it b}) Numerical solution for $h$ at $t=3$.
({\it c}) Magnitude of the derivative $u_x$ of the solution for the scheme~\eqref{eq:InvariantSchemeShallowWaterComputationalCoordinates}. Light shades refer to high values of $|u_x|$.
}\label{fig:ImplicitInvariantSchemeOneDimensionalShallowWaterResult}
\end{figure}

In Fig.~\ref{fig:ImplicitInvariantSchemeOneDimensionalShallowWaterResult} we show the integration of the one-dimensional shallow-water equations using the scheme~\eqref{eq:InvariantSchemeShallowWaterComputationalCoordinates}, \eqref{eq:DiscretizationEquidistributionPrinciple} with arc-length monitor function discretization~\eqref{eq:DiscretizationInvariantMonitorFunction} utilizing the same initial conditions as those chosen for the integration shown in Fig.~\ref{fig:LagrangianImplicitInvariantSchemeOneDimensionalShallowWaterResult}. It is clearly visible that the mesh points remain almost fixed as long as the shock is not developed. Once the shock is traveling through the domain, the mesh points are able to sufficiently adapt to yield increased resolution in the region near the shock (as additionally shown in Fig.~\ref{fig:ImplicitInvariantSchemeOneDimensionalShallowWaterResult}{\it c}). Again the scheme approximately conserves mass and momentum but dissipates energy. The relative errors in the momentum and energy conservation are approximately the same as in the case of the Lagrangian schemes in the previous subsection.

It is worth pointing out that the time step of the integration shown in Fig.~\ref{fig:ImplicitInvariantSchemeOneDimensionalShallowWaterResult} is relatively small. The reason for this is that by using the scheme~\eqref{eq:InvariantSchemeShallowWaterComputationalCoordinates} and~\eqref{eq:DiscretizationEquidistributionPrinciple}--\eqref{eq:DiscretizationInvariantMonitorFunction} we decouple the solution of the physical differential equation and the equation controlling the location of grid points. If time steps were not small, a severe time lag in the mesh movement would occur and the resulting mesh would not satisfy the equidistribution principle closely enough to give a satisfactory adaptivity. The above problem was extensively addressed in~\cite{huan10By}. It can be overcome via the iterative solution of the physical and mesh equations a number of times, which leads to a reduction of the time lag in the mesh movement. Such a strategy could be readily adopted with the scheme~\eqref{eq:InvariantSchemeShallowWaterComputationalCoordinates} and~\eqref{eq:DiscretizationEquidistributionPrinciple}--\eqref{eq:DiscretizationInvariantMonitorFunction} because a repeated iterative integration does not break the invariance of this scheme.

In order to facilitate the comparison of distinct types of invariant numerical schemes, we also keep the time step small in the integration of the Lagrangian scheme for the one-dimensional shallow-water equations, which is presented in Fig.~\ref{fig:LagrangianImplicitInvariantSchemeOneDimensionalShallowWaterResult}.

\section{Invariant numerical models for the two-dimensional\texorpdfstring{\\}{ }shallow-water equations}\label{sec:TwoDimensionalShallowWater}

\subsection{Selection of symmetries using boundary conditions}\label{sec:2DShallowWaterSymmetriesOfBVE}

The domains most often considered in geophysical fluid dynamics for the numerical integration of the two-dimensional shallow-water equations on a plane are either a channel with periodic boundary conditions in the East--West direction and rigid boundaries in the North--South direction or a domain with double periodic boundary conditions. As the second configuration is more challenging from the point of view of invariant numerical schemes, we will employ it subsequently.

Lie symmetry operators of the two-dimensional shallow-water equations~\eqref{eq:TwoDimensionalShallowWater} with periodic boundary conditions in both the East--West and North--South directions form the five-dimen\-sional subalgebra~$\mathfrak s_2$ of the maximal Lie invariance algebra~$\mathfrak g_2$ of the equations~\eqref{eq:TwoDimensionalShallowWater} without additional constraints.
A basis of~$\mathfrak s_2$ is given by
\begin{equation}
\label{eq:SymmetriesDoublePeriodicDomain}
    \p_t,\quad\p_x,\quad\p_y,\quad t\p_x+\p_u,\quad t\p_y+\p_v.
\end{equation}
As in Section~\ref{sec:OneDimensionalShallowWater} we could additionally include the scaling symmetries of the equations~\eqref{eq:TwoDimensionalShallowWater} in the  subalgebra~$\mathfrak s_2$, referring to them as equivalence transformations of the class of doubly periodic boundary value problems. The reason why we did not include these scalings above is that all the discretizations for the shallow-water equations we use subsequently do not change the scaling properties of that system. This means that these discretizations already satisfy the required scaling properties by construction. On the other hand, the additional presence of scaling operators would slightly complicate the expressions for the difference invariants computed below, without giving any significant new information (the additional coefficients arising will factor out anyway for the resulting schemes). Only in the course of setting up the invariant grid generator will it be necessary to explicitly take into account the specific scaling symmetries, which will consistently be done in Section~\ref{sec:TwoDimensionalShallowWaterEulerian}. Note that both symmetries~\eqref{eq:SymmetriesDoublePeriodicDomain} and scaling symmetries of the two-dimensional shallow-water equations~\eqref{eq:TwoDimensionalShallowWater} generate equivalence transformations of the set of relevant initial conditions.

As can be envisioned from the consideration of the numerical models of the one-dimensional shallow-water equations discussed in Section~\ref{sec:OneDimensionalShallowWater},  discretization schemes for the two-dimensional shallow-water equations will be invariant under Galilean symmetries only if they are based on adaptive grids. Because for the channel model Galilean transformations are only admitted in the $x$-direction, it suffices for a grid to be adaptive in the $x$-direction. This in particular means that we can use a uniform spacing in the $y$-direction and have a spatial grid with changing resolution only along the channel. On the other hand, the shallow-water equations with double periodic boundary conditions require the treatment of adaptive grids in both the $x$- and $y$-directions. An initial orthogonal spatial grid is driven to a nonorthogonal grid, which makes the direct evaluation of finite difference derivatives much more elaborate. This problem is treated upon using a finite volume formulation of that scheme.

For simplicity, all the schemes are developed on an Arakawa A-grid subsequently; i.e., the variables $u$, $v$, $h$ are defined in the same respective points. See, e.g.,~\cite{rand94Ay} for a discussion of different types of staggered grids for the shallow-water equations. Other types of grid staggering can be used in a similar way as that shown for the A-grid in what follows.

\subsection{Invariant numerical schemes with double periodic boundary conditions:\texorpdfstring{\\}{ }Lagrangian scheme}\label{sec:TwoDimensionalShallowWaterLagrangian}

The main difficulty with adaptive grids in both the $x$- and $y$-directions is that it can become cumbersome to directly evaluate the gradients of the dependent variables on such curvilinear grids by finite differences. As discussed in Section~\ref{sec:OneDimensionalShallowWater}, a prominent strategy to overcome this difficulty is to introduce a mapping from the computational (logical) coordinates $(\xi,\eta)$ to the physical coordinates $(x,y)$. This will be done in Section~\ref{sec:TwoDimensionalShallowWaterEulerian}.

For the sake of demonstration we take another, more direct approach here, namely using the finite volume formulation of the divergence operator; see, e.g.,~\cite{ring02Ay}. Using the theorem of Gau\ss--Ostrogradsky, we can approximate the divergence $\nn\cdot\mathbf f$ of a vector-function~$\mathbf f$ over a single grid cell with area $A$ and edge lengths $l_i$ as
\[
 \nn\cdot\mathbf f \approx \frac{1}{A}\sum_{i=1}^4(\mathbf{f}_i\cdot\mathbf{n}_i)l_i.
\]
In the above formula, $\mathbf{n}_i$ denotes the outward-directed unit vector at the single cell edges.

As it is possible to cast the shallow-water equations~\eqref{eq:TwoDimensionalShallowWater} into conserved (momentum) form,
\begin{align}\label{eq:ShallowWaterConservedForm}
\begin{split}
& {\rm Eq}^h = h_t + (hu)_x+(hv)_y = 0, \\
& {\rm Eq}^u = (hu)_t + \left(hu^2+\frac12 h^2\right)_x+(huv)_y=0, \\
& {\rm Eq}^v = (hv)_t + (huv)_x+ \left(hv^2+\frac12 h^2\right)_y=0,
\end{split}
\end{align}
the above approximation of the divergence operator is sufficient to discretize the two-dimensional shallow-water equations using the finite volume form.\footnote{Equations including curl terms can be converted into finite volume representation using the Stokes theorem.} On the other hand, a finite volume discretization is readily applicable on adaptive grids, as it is not necessary to approximate derivatives by finite differences in such a formulation.

In order to discretize~\eqref{eq:ShallowWaterConservedForm} in an invariant way using the Dorodnitsyn method, we would need a set of difference invariants and would need to construct the discretization using these invariants as building blocks for the numerical scheme. The problem with this approach is the same as reported in the one-dimensional case. The Galilean transformation $\tilde t=t$, $\tilde x=x+\varepsilon_1t$, $\tilde y=y+\varepsilon_2t$,
$\tilde h=h$, $\tilde u=u+\varepsilon_1$, $\tilde v=v+\varepsilon_2$ maps the system~\eqref{eq:ShallowWaterConservedForm} to
\begin{equation}\label{eq:GalileanTransformation2d}
\widetilde{\rm Eq}{}^h={\rm Eq}^h,\quad
\widetilde{\rm Eq}{}^u={\rm Eq}^u+\varepsilon_1{\rm Eq}^h,\quad
\widetilde{\rm Eq}{}^v={\rm Eq}^v+\varepsilon_2{\rm Eq}^h;
\end{equation}
i.e., it leads to a combination of the momentum equations with the continuity equation. Expressing the momentum equations in terms of differential invariants thus again only works by combining these equations with the continuity equation. As in the one-dimensional case, it is therefore not natural to attempt to find an invariant approximation of the momentum form of the shallow-water equations using difference invariants.

At the same time, the Lagrangian form of the shallow-water equations~\eqref{eq:TwoDimensionalShallowWater}, which is
\begin{gather}\label{eq:TwoDimensionalShallowWaterLagrangian}
 \dd{x}{t}=u,\quad \dd{y}{t}=v,\quad
 \dd{h}{t}+h\left(\pdl{u}{x}+\pdl{v}{y}\right)=0,\quad \dd{u}{t}+\pdl{h}{x}=0,\quad \dd{v}{t}+\pdl{h}{y}=0,
\end{gather}
can be approximated using the finite volume method as well. Expressing an invariant discretization of~\eqref{eq:TwoDimensionalShallowWaterLagrangian} in terms of difference invariants is considerably easier than doing the same for an invariant discretization of~\eqref{eq:ShallowWaterConservedForm}. As in the one-dimensional case, the drawback of using~\eqref{eq:TwoDimensionalShallowWaterLagrangian} as a starting point is that the resulting scheme does not approximate a conserved form and thus preserves neither mass and momenta nor energy.

\begin{figure}[ht!]
\centering
\includegraphics[scale=0.98]{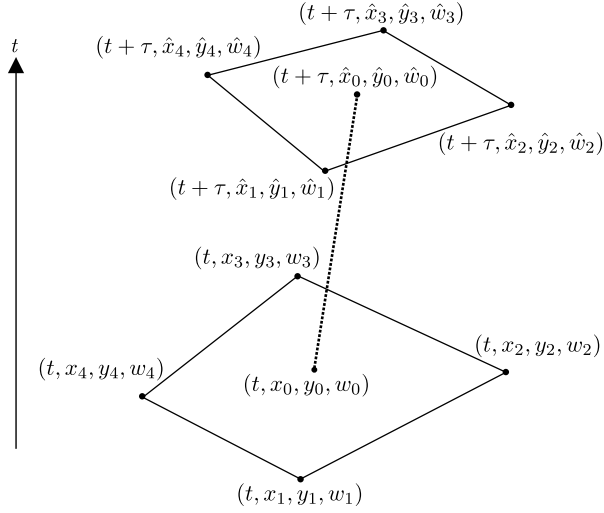}
\caption{Stencil for the invariant Lagrangian schemes for the two-dimensional shallow-water equations with double periodic boundary conditions. The dependent variables $w=(u,v,h)$ are defined in the center $(x_0,y_0)$ of the cells of the respective polygons. The fluxes are computed using the values at the corners $(x_k,y_k), k=1\dots 4$, of the cells, which are obtained by interpolation from the values at the polygon centroids. Variables with a hat are those at the subsequent time step.}
\label{fig:StencilTwoDim}
\end{figure}

The stencil of the discretizations we aim to use is given in Fig.~\ref{fig:StencilTwoDim}. All the dependent variables are defined in the centroids of the respective polygons. The fluxes through the edges will govern the evolution of these centroid values. In order to facilitate the computation of the fluxes it is necessary to determine the values of $w=(u,v,h)$ in the cell corners, which is done by interpolation. While in principle any type of interpolation can be used, we employ natural neighbors interpolation for this purpose, i.e., the values at the cell corners are
$
 w_j= \sum_{\kappa=1}^4\rho^{\kappa,j} w^{\kappa,j}_0,
$
where $j=1,\dots,4$ and $w^{\kappa,j}_0$ are the values of~$w$ in the centers of those cells having in common the corner denoted by~$j$. The interpolation weights~$\rho^{\kappa,j}$ are determined in the following way. The Voronoi tessellation generated by the cell centers is constructed. Then a new tessellation is computed in which the point $(x_j,y_j)$ is introduced as an additional generator. Denote by $A_{P^\kappa_0}$ the area of the Voronoi cell of the original tessellation associated with the center point $P^\kappa_0=(x^\kappa_0,y^\kappa_0)$ and by $A_{P_j}$ the area of the new cell associated with the corner point $P_j=(x_j,y_j)$ introduced for the second tessellation. 
Then the weights~$\rho^{\kappa,j}$ are computed as $\rho^{\kappa,j}= (A_{P_j} \cap A_{P^\kappa_0})/A_{P_j}$. Once the values $w_j$ are obtained, they can be regarded as proper stencil variables. 

The prolongations of the symmetry operators~\eqref{eq:SymmetriesDoublePeriodicDomain} on the variables of the stencil shown in Fig.~\ref{fig:StencilTwoDim} read
\begin{align}\label{eq:DifferenceInvariantsTwoDimensional}
\begin{split}
    &\p_t, \quad \sum_{i=0}^{4}(\p_{x_i}+\p_{\hat x_i}), \quad \sum_{i=0}^{4}(\p_{y_i}+\p_{\hat y_i}),\\
    &\sum_{i=0}^{4}(t\p_{x_i}+(t+\tau)\p_{\hat x_i}+\p_{u_i}+\p_{\hat u_i}),\quad
    \sum_{i=0}^{4}(t\p_{y_i}+(t+\tau)\p_{\hat y_i}+\p_{v_i}+\p_{\hat v_i}).
\end{split}
\end{align}
These prolongations are well agreed with the above interpolation procedure.
The difference invariants of the set~\eqref{eq:DifferenceInvariantsTwoDimensional} are given by
\begin{align*}
  &\tau,\quad  h_i,\quad \hat h_i,\quad
  x_i-x_j,\quad y_i-y_j,\quad \hat x_i-x_j-\tau u_k,\quad \hat y_i-y_j-\tau v_k,\\
  &u_i-u_j,\quad \hat u_i-u_j,\quad v_i-v_j,\quad \hat v_i-v_j,
\end{align*}
where the indices~$i$, $j$ and~$k$ take the values $0,\dots,4$. Note that of course not all of the above difference invariants are independent if $i$, $j$ and~$k$ separately run through all possible values.

A simple explicit invariant scheme (Euler forward scheme) that can be constructed using these invariants is
\begin{align}\label{eq:ExplicitInvariantSchemeTwoDimensionalShallowWater}
\begin{split}
 &\frac{\hat x_0-x_0}{\tau} - u_0 = 0,\quad \frac{\hat y_0-y_0}{\tau} - v_0 = 0,\\
 &\frac{\hat h_0 - h_0}{\tau} +  \frac{h_0}{2A}\sum_{i=1}^4[(u_i+u_{i+1})(y_{i+1}-y_i)- (v_i+v_{i+1})(x_{i+1}-x_i)]=0,\\
 &\frac{\hat u_0 - u_0}{\tau} + \frac{1}{2A}\sum_{i=1}^4(h_i+h_{i+1})(y_{i+1}-y_{i})=0,\\
 &\frac{\hat v_0 - v_0}{\tau} - \frac{1}{2A}\sum_{i=1}^4(h_i+h_{i+1})(x_{i+1}-x_{i})=0,
\end{split}
\end{align}
where $A=\frac12\sum_{i=1}^4(x_iy_{i+1}-x_{i+1}y_i)$ is the area of the polygon spanned by $(x_1,y_1), \dots, (x_4,y_4)$ and $(x_5,y_5,u_5,v_5,h_5)=(x_1,y_1,u_1,v_1,h_1)$ by definition. As in the one-dimensional case, in the continuous limit this scheme converges to the Lagrangian representation of the two-dimensional shallow-water equations~\eqref{eq:TwoDimensionalShallowWaterLagrangian}.

\begin{figure}[t!]
\centering
\includegraphics{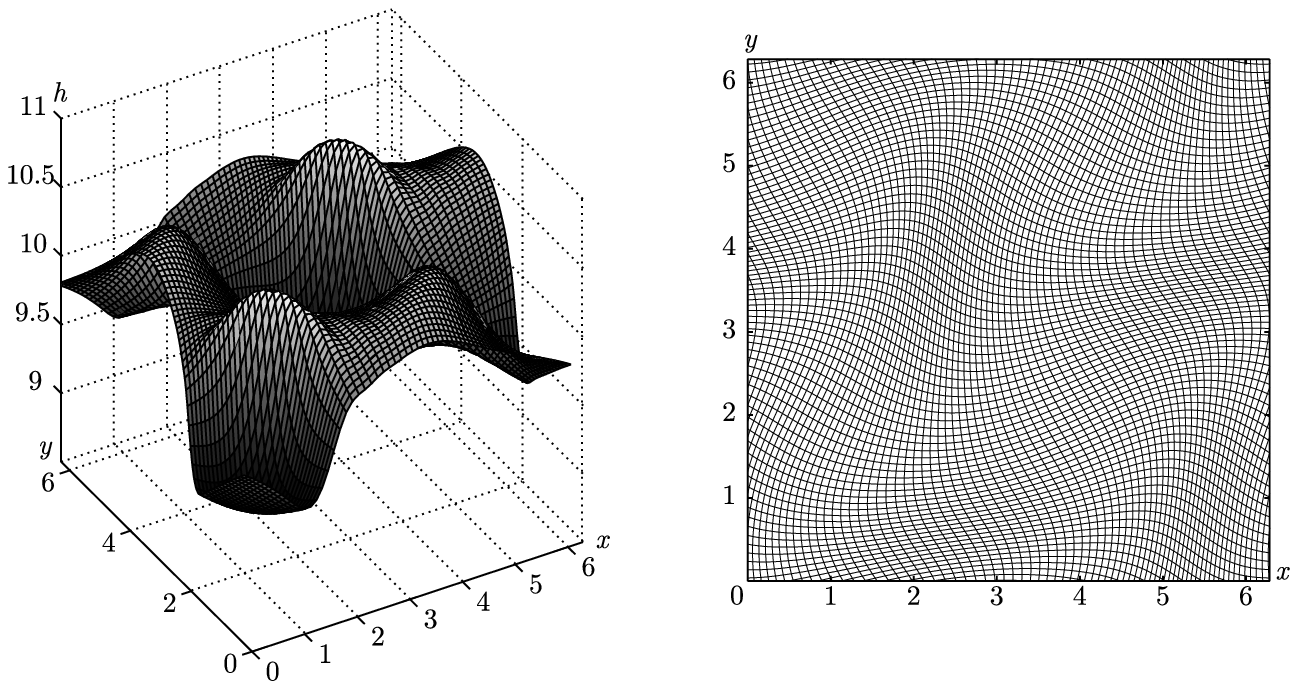}
\caption{Numerical integration of the two-dimensional shallow-water equations~\eqref{eq:TwoDimensionalShallowWater} using the scheme~\eqref{eq:ImplicitInvariantSchemeTwoDimensionalShallowWater} with $\tau=0.001$ and $N_x\times N_y=71\times71$ grid points on the square $[0, 2\pi]\times [0, 2\pi]$ over the time interval $[0,2]$. The initial conditions are $u=A\sin(x+\varphi_0)\sin y$, $v=A\sin x\sin y$ and $h=h_0+A\cos(x+\varphi_0)\cos y$, with $A=0.4$, $\varphi_0=\pi/6$ and $h_0=10$. Left: Numerical solution for $h$ at $t=2$. Right: Spatial discretization grid at $t=2$.}
\label{fig:ImplicitInvariantSchemeTwoDimensionalShallowWaterResult}
\end{figure}

In a similar manner, we can formulate the implicit scheme (trapezoidal rule)
\begin{align}\label{eq:ImplicitInvariantSchemeTwoDimensionalShallowWater}
\begin{split}
 \frac{\hat x_0-x_0}{\tau} - {}
 &\frac12(u_0+\hat u_0) = 0,\quad \frac{\hat y_0-y_0}{\tau} - \frac12(v_0+\hat v_0) = 0,\\
 \frac{\hat h_0 - h_0}{\tau} + {}
 &\frac{h_0}{4A}\sum_{i=1}^4[(u_i+u_{i+1})(y_{i+1}-y_i)- (v_i+v_{i+1})(x_{i+1}-x_i)]+{}\\
 &{} \frac{\hat h_0}{4\hat A}\sum_{i=1}^4[(\hat u_i+\hat u_{i+1})(\hat y_{i+1}-\hat y_i)- (\hat v_i+\hat v_{i+1})(\hat x_{i+1}-\hat x_i)]=0,\\
 \frac{\hat u_0-u_0}{\tau} + {}
 &\frac{1}{4A}\sum_{i=1}^4(h_i+h_{i+1})(y_{i+1}-y_{i})+\frac{1}{4\hat A}\sum_{i=1}^4(\hat h_i+\hat h_{i+1})(\hat y_{i+1}-\hat y_{i})=0,\\
 \frac{\hat v_0 - v_0}{\tau} - {}
 &\frac{1}{4A}\sum_{i=1}^4(h_i+h_{i+1})(x_{i+1}-x_{i})- \frac{1}{4\hat A}\sum_{i=1}^4(\hat h_i+\hat h_{i+1})(\hat x_{i+1}-\hat x_{i})=0.
\end{split}
\end{align}
Fig.~\ref{fig:ImplicitInvariantSchemeTwoDimensionalShallowWaterResult} shows the result of a numerical integration with the scheme~\eqref{eq:ImplicitInvariantSchemeTwoDimensionalShallowWater} supplemented with periodic boundary conditions and specific initial conditions. The numerical solution of the water height~$h$ at~$t=2$ is shown in the left panel. The right panel depicts the associated discretization grid at $t=2$. As in the case of the one-dimensional Lagrangian scheme, a strong distortion of the grid cells is visible, which is not directly related to pronounced features in the numerical solution, but rather a consequence of the Lagrangian grid movement. As both the discretizations~\eqref{eq:ExplicitInvariantSchemeTwoDimensionalShallowWater} and~\eqref{eq:ImplicitInvariantSchemeTwoDimensionalShallowWater} do not approximate the conserved form of the shallow-water equations~\eqref{eq:ShallowWaterConservedForm}, they conserve neither the mass~$\mathcal M$ and the momenta~$\mathcal P_x$ and $\mathcal P_y$ nor the energy~$\mathcal H$.
\looseness=1


It should be stressed that it is possible to formulate an invariant finite volume scheme for the conserved form of the shallow-water equations~\eqref{eq:ShallowWaterConservedForm}, in a similar way as was shown in the previous section for the one-dimensional case. As said above, the problem of doing this systematically within the Dorodnitsyn approach is that it can be hard to find a proper combination of elementary difference invariants that allows one to approximate the momentum form of the shallow-water equations. This is why we will show an alternative way of constructing invariant numerical schemes for the two-dimensional shallow-water equations in the following section, which will avoid the technical complications that can arise when using the difference invariants method for the construction of symmetry-preserving numerical schemes.

\subsection{Invariant numerical schemes with double periodic boundary conditions:\texorpdfstring{\\}{ }Eulerian scheme}\label{sec:TwoDimensionalShallowWaterEulerian}

Though the finite volume discretization developed in Section~\ref{sec:TwoDimensionalShallowWaterLagrangian} is suitable from the point of view of invariance preservation, it is not ideal from the viewpoint of numerical analysis. In general, Lagrangian schemes are not in widespread use as they can easily lead to tangling meshes or rapidly changing grids through the spatial domain. For the same reason, numerical schemes in hydrodynamics are usually formulated in terms of Eulerian variables (or using some combination of Eulerian and Lagrangian schemes).
An invariant scheme on an adaptive grid can be formulated by combining the idea of having an invariant grid generator proposed in Section~\ref{sec:AlternativeSolutionProcedure} with the discretization in computational coordinates.
More specifically, we consider the momentum form~\eqref{eq:ShallowWaterConservedForm} of the two-dimensional shallow-water equations and rewrite it in the computational coordinates $\theta=t$, $\xi=\xi(t,x,y)$, $\eta=\eta(t,x,y)$:
\begin{equation}\label{eq:2DShallowWaterCompCoords}
 \pdl{}{\theta}{(JF^t)}+\pdl{}{\xi}(J\xi_tF^t+J\xi_x F^x+J\xi_y F^y)+\pdl{}{\eta}(J\eta_tF^t+J\eta_x F^x+J\eta_y F^y)=0,
\end{equation}
where $F^t=(h,hu,hv)$, $F^x=(hu,hu^2+\frac12h^2,huv)$, $F^y=(hv,huv,hv^2+\frac12h^2)$ and
\begin{gather*}
J=x_\xi y_\eta-x_\eta y_\xi,\quad \xi_t = -\xi_x x_\theta - \xi_y y_\theta,\quad \eta_t = -\eta_x x_\theta-\eta_y y_\theta,\\
\xi_x = \frac{y_\eta}J,\quad \xi_y = -\frac{x_\eta}J, \quad \eta_x=-\frac{y_\xi}J,\quad \eta_y=\frac{x_\xi}J.
\end{gather*}
The invariance of the system~\eqref{eq:2DShallowWaterCompCoords}
with respect to shifts of~the former independent variables $t$, $x$ and~$y$
is obvious since the left-hand sides of equations of the system
(which we denote by ${\rm Eq}^h$, ${\rm Eq}^u$ and ${\rm Eq}^v$, respectively)
do not explicitly involve these variables.
Therefore, any finite difference approximation of~\eqref{eq:2DShallowWaterCompCoords} is invariant with respect to the above shifts
extended to the corresponding stencil; cf.~\eqref{eq:DifferenceInvariantsTwoDimensional}.
Note that any involved transformation is trivially extended to the computational coordinates~$\xi$ and~$\eta$; i.e., they are not transformed.
The scale symmetry transformations of the shallow-water equations~\eqref{eq:TwoDimensionalShallowWater}
are automatically preserved in the course of a proper finite difference approximation of~\eqref{eq:2DShallowWaterCompCoords};
see the related discussion in Section~\ref{sec:2DShallowWaterSymmetriesOfBVE}.
In order to make clear the invariance with respect to Galilean boosts, we recombine terms in~\eqref{eq:2DShallowWaterCompCoords},
substituting $J\xi_t=-J\xi_x x_\theta - J\xi_y y_\theta$ and $J\eta_t=-J\eta_x x_\theta-J\eta_y y_\theta$:
\begin{align*}
\begin{split}
 \pdl{}{\theta}{(JF^t)}+{}
 &\pdl{}{\xi}\big(J\xi_x(F^x-x_\theta F^t)+J\xi_y(F^y-y_\theta F^t)\big)+{}\\
 &\pdl{}{\eta}\big(J\eta_x(F^x-x_\theta F^t)+J\eta_y(F^y-y_\theta F^t)\big)=0.
\end{split}
\end{align*}
The Galilean boost
$\tilde t=t$, $\tilde x=x+\varepsilon_1t$, $\tilde y=y+\varepsilon_2t$,
$\tilde h=h$, $\tilde u=u+\varepsilon_1$, $\tilde v=v+\varepsilon_2$
maps the system~\eqref{eq:2DShallowWaterCompCoords} to the system with
\[
\widetilde{\rm Eq}{}^h={\rm Eq}^h,\quad
\widetilde{\rm Eq}{}^u={\rm Eq}^u+\varepsilon_1{\rm Eq}^h,\quad
\widetilde{\rm Eq}{}^v={\rm Eq}^v+\varepsilon_2{\rm Eq}^h.
\]
Note that this is the same transformation law in computational variables as it is in the physical space~\eqref{eq:GalileanTransformation2d}. \emph{The main idea for finding invariant numerical schemes of~\eqref{eq:2DShallowWaterCompCoords} is to construct the discretization in such a manner that the discrete counterpart of the system~\eqref{eq:2DShallowWaterCompCoords} is transformed similarly
by the extension of the Galilean boost to the stencil points.}
In order to preserve Galilean boosts as symmetries in the course of discretization, it suffices
\begin{itemize}\itemsep=0ex
\item
to use the same discretization schemes for all the equations of the system~\eqref{eq:2DShallowWaterCompCoords},
just running through the number of corresponding components of $F^t$, $F^x$ and $F^y$;
\item
to evaluate all the copies of $J\xi_x$ (resp.\ the components~$F^t$ and~$F^x$) related to the block $J\xi_x(F^x-x_\theta F^t)$
in the same grid point and in the same way; the same rule should be applied for the other similar blocks:
$J\xi_y(F^y-y_\theta F^t)$, $J\eta_x(F^x-x_\theta F^t)$ and $J\eta_y(F^y-y_\theta F^t)$.
\end{itemize}
For example, consider the trapezoidal scheme for the system~\eqref{eq:2DShallowWaterCompCoords}:
\begin{equation}\label{eq:2DShallowWaterCompCoordsScheme}
 \frac{\hat J_{jk}\hat F^t_{jk}-J_{jk}F^t_{jk}}\tau+\frac{U_{jk}+\hat U_{jk}}2+\frac{V_{jk}+\hat V_{jk}}2=0,
\end{equation}
where $\tau$ is the step in $\theta=t$;
\begin{align*}
 U_{jk}=\smash{\frac{1}{2\Delta\xi}}[
 &(J\xi_t)_{j+1/2,k}(F^t_{jk}+F^t_{j+1,k})-(J\xi_t)_{j-1/2,k}(F^t_{jk}+F^t_{j-1,k})+{}\\
 &(J\xi_x)_{j+1/2,k}(F^x_{jk}+F^x_{j+1,k})-(J\xi_x)_{j-1/2,k}(F^x_{jk}+F^x_{j-1,k})+{}\\
 &(J\xi_y)_{j+1/2,k}(F^y_{jk}+F^y_{j+1,k})-(J\xi_y)_{j-1/2,k}(F^y_{jk}+F^y_{j-1,k})],\\[1ex]
 V_{jk}=\smash{\frac{1}{2\Delta\eta}}[
 &(J\eta_t)_{j,k+1/2}(F^t_{jk}+F^t_{j,k+1})-(J\eta_t)_{j,k-1/2}(F^t_{jk}+F^t_{j,k-1})+{}\\
 &(J\eta_x)_{j,k+1/2}(F^x_{jk}+F^x_{j,k+1})-(J\eta_x)_{j,k-1/2}(F^x_{jk}+F^x_{j,k-1})+{}\\
 &(J\eta_y)_{j,k+1/2}(F^y_{jk}+F^y_{j,k+1})-(J\eta_y)_{j,k-1/2}(F^y_{jk}+F^y_{j,k-1})];
\end{align*}
the values $J$, $J\xi_x=y_\eta$, $J\xi_y=-x_\eta$, $J\eta_x=-y_\xi$, $J\eta_y=x_\xi$, $J\xi_t$ and $J\eta_t$
are discretized in the following way:
\begin{align*}
 &J_{jk}=\frac1{4\Delta\xi\Delta\eta}[(x_{j+1,k}-x_{j-1,k})(y_{j,k+1}-y_{j,k-1})-(x_{j,k+1}-x_{j,k-1})(y_{j+1,k}-y_{j-1,k})],\\
 &(J\xi_x)_{j\pm1/2,k}=\frac1{4\Delta\eta}(y_{j,k+1}-y_{j,k-1}+y_{j\pm1,k+1}-y_{j\pm1,k-1}),\\
 &(J\xi_y)_{j\pm1/2,k}=-\frac1{4\Delta\eta}(x_{j,k+1}-x_{j,k-1}+x_{j\pm1,k+1}-x_{j\pm1,k-1}),\\
 &(J\eta_x)_{j,k\pm1/2}=-\frac1{4\Delta\xi}(y_{j+1,k}-y_{j-1,k}+y_{j+1,k\pm1}-y_{j-1,k\pm1}),\\
 &(J\eta_y)_{j,k\pm1/2}=\frac1{4\Delta\xi}(x_{j+1,k}-x_{j-1,k}+x_{j+1,k\pm1}-x_{j-1,k\pm1}),\\
 &(J\xi_t)_{j\pm1/2,k} = -(J\xi_x)_{j\pm1/2,k}\frac{\dot x_{jk}+\dot x_{j\pm1,k}}2-(J\xi_y)_{j\pm1/2,k}\frac{\dot y_{jk}+\dot y_{j\pm1,k}}2,\\
 &(J\eta_t)_{j,k\pm1/2} = -(J\eta_x)_{j,k\pm1/2}\frac{\dot x_{jk}+\dot x_{j,k\pm1}}2-(J\eta_y)_{j,k\pm1/2}\frac{\dot y_{jk}+\dot y_{j,k\pm1}}2,
\end{align*}
where $\dot x_{jk}= (\hat x_{jk}-x_{jk})/\tau$ and $\dot y_{jk}= (\hat y_{jk}-y_{jk})/\tau$ are by definition the mesh velocities in the $x$- and $y$-directions, respectively;
and by hat we mark the corresponding values at the time $\theta+\tau$. In particular, we take
\begin{align*}
 &\hat{(J\xi_t)}_{j\pm1/2,k} = -\hat{(J\xi_x)}_{j\pm1/2,k}\frac{\dot x_{jk}+\dot x_{j\pm1,k}}2
 -\hat{(J\xi_y)}_{j\pm1/2,k}\frac{\dot y_{jk}+\dot y_{j\pm1,k}}2,\\
 &\hat{(J\eta_t)}_{j,k\pm1/2} = -\hat{(J\eta_x)}_{j,k\pm1/2}\frac{\dot x_{jk}+\dot x_{j,k\pm1}}2
 -\hat{(J\eta_y)}_{j,k\pm1/2}\frac{\dot y_{jk}+\dot y_{j,k\pm1}}2.
\end{align*}

As the system of difference equations~\eqref{eq:2DShallowWaterCompCoordsScheme} satisfies the above conditions,
it is invariant with respect to properly extended Galilean boosts.

\begin{remark}
 The usage of computational coordinates also underlines the subtle change of the meaning of the time derivatives in a number of papers devoted to the construction of invariant numerical schemes, such as in~\cite{doro00Ay,vali05Ay}. While in the standard (Eulerian) discretization, the continuous limit of the form $(\hat u-u)/\tau$ yields the partial derivative $u_t$, in the framework of invariant schemes these terms are often to be interpreted as Lagrangian time derivatives $\dot u$ (see also Section~\ref{sec:ClassicalInvariantNumericalSchemes}). This immediate transition from an Eulerian (partial) time derivative to a Lagrangian (total) time derivative is a necessary consequence of the intermediate step of discretizing an equation in computational coordinates and assuming that the grid evolution is described by the equations $\dot x=u$ and $\dot y=v$; see also the discussion in~\cite{reic99Ay}.
\end{remark}

\begin{remark}
It should be noted that computational coordinates have a clear physical meaning in the present context. As they do not change during the evolution of the grid, they can be interpreted as the Lagrangian variables (fluid labels) of fluid mechanics provided we again assume a Lagrangian grid evolution. A prominent way to choose these fluid labels is by setting them to equal the Cartesian coordinates at the onset of evolution. By definition, this is the same role that computational coordinates play in the numerics of moving meshes. Stated in another way, the requirement of maintaining invariance of the discretization scheme and discretization stencil of the shallow-water equations under the Galilean group naturally boils down to discretizing these equations in computational coordinates.
\end{remark}

It then remains to specify the grid velocities $\dot x$ and $\dot y$ in order to complete the scheme given in~\eqref{eq:2DShallowWaterCompCoordsScheme}. This can be done in a similar manner as in Section~\ref{sec:AlternativeSolutionProcedure} using the idea of equidistributing meshes (though, strictly speaking, equidistribution in higher dimensions is not sufficient to uniquely determine an adaptive grid; see, e.g., the discussion in~\cite{huan10By}). Thus, the grid will be determined from the system of elliptic equations
\begin{equation}\label{eq:GridGeneratorTwoDimensionalContinuous}
 \nn_{\boldsymbol{\xi}}\cdot (\mathbb{G}\nn_{\boldsymbol{\xi}}x) = 0,\quad \nn_{\boldsymbol{\xi}}\cdot (\mathbb{G}\nn_{\boldsymbol{\xi}}y) = 0.
\end{equation}
Here $\nn_{\boldsymbol{\xi}}$ denotes the gradient in the space of computational coordinates $(\xi,\eta)$ and $\mathbb{G}=w\mathbb{I}$ is the matrix-valued monitor function, where $\mathbb{I}$ is the two-by-two unit matrix and $w=w(x,y)$ is a weight function which depends on the (numerical) solution of the shallow-water equations~\cite{cao99Ay}. An invariant discretization of~\eqref{eq:GridGeneratorTwoDimensionalContinuous} reads
\begin{align*}
 &\frac{1}{\Delta\xi}\left(w_{i+1/2,j}\frac{\hat z_{i+1,j}-\hat z_{ij}}{\Delta\xi}-w_{i-1/2,j}\frac{\hat z_{ij}-\hat z_{i-1,j}}{\Delta\xi}\right)+{}\\
 &{}\frac{1}{\Delta\eta}\left(w_{i,j+1/2}\frac{\hat z_{i,j+1}-\hat z_{ij}}{\Delta\eta}-w_{i,j-1/2}\frac{\hat z_{ij}-\hat z_{i,j-1}}{\Delta\eta}\right)=0,
\end{align*}
where $z=x$ and $z=y$ for the first and second equations, respectively, and
\begin{gather*}
w_{i+1/2,j}=\frac{w_{i+1,j}+w_{ij}}2,\quad w_{i-1/2,j}=\frac{w_{ij}+w_{i-1,j}}2,\\
w_{i,j+1/2}=\frac{w_{i,j+1}+w_{ij}}2,\quad w_{i,j-1/2}=\frac{w_{ij}+w_{i,j-1}}2
\end{gather*}
provided that~$w$ is approximated by a difference invariant of the algebra~$\mathfrak s_2$, which is spanned by the vector fields~\eqref{eq:SymmetriesDoublePeriodicDomain}. Once again, straightforward choices for~$w$ that can be discretized using difference invariants are
\begin{align*}
 w^1=\sqrt{1+\alpha(u_x^2+u_y^2+v_x^2+v_y^2)},\qquad
 w^2=\sqrt{1+\alpha(h_{xx}+h_{yy})^2},
\end{align*}
but of course other forms for~$w$ are possible as well.
Similarly to the one-dimensional case (cf.\ Remark~\ref{re:OnChoice ofInvMonitorFunctions1D}),
the general form of~$w$ that is an invariant of the algebra~$\mathfrak s_2$ is given by an arbitrary smooth function of derivatives of~$u$, $v$ and~$h$ with respect to~$x$ and~$y$ including $h$ itself but not~$u$ and~$v$. At the same time, we should also take into account other desired properties of~$w$.
Both the functions~$w^1$ and~$w^2$
are invariant with respect to shifts and Galilean boosts generated by vector fields~\eqref{eq:SymmetriesDoublePeriodicDomain},
the scalings generated by the vector field $x\p_x+y\p_y+u\p_u+v\p_v+2h\p_h$ and even rotations.
All the scaling symmetries of the shallow-water equations
are at least equivalence transformations for the sets of functions of such forms,
where the parameter~$\alpha$ is varied.

It should also be stressed that the grid generator based on system~\eqref{eq:GridGeneratorTwoDimensionalContinuous} is a rather simple one. More advanced formulations of grid generators exist, e.g., by using a general positive definite and symmetric matrix~$\mathbb{G}$. Alternatively, the grids at a certain time level could be computed using moving mesh partial differential equations~\cite{budd09Ay,huan10By}, provided it is possible to discretize such equations in an invariant way.

A different methodology is to use so-called \textit{velocity-based} moving mesh strategies. Unlike the \textit{location-based} methods, which were exclusively used in the present paper, in the velocity-based methods the location of the grid points is not determined directly but rather equations for the mesh velocity are formulated.
Velocity based strategies, such as the method involving the geometric conservation law~\cite{cao02Ay,huan10By}, provide alternative ways of formulating grid equations that give a basis for realizing invariant moving mesh equations. See also~\cite{thom97Ay}, where the term \emph{geometric conservation law} was introduced.

\begin{figure}[ht!]
\centering
\includegraphics[scale=0.97]{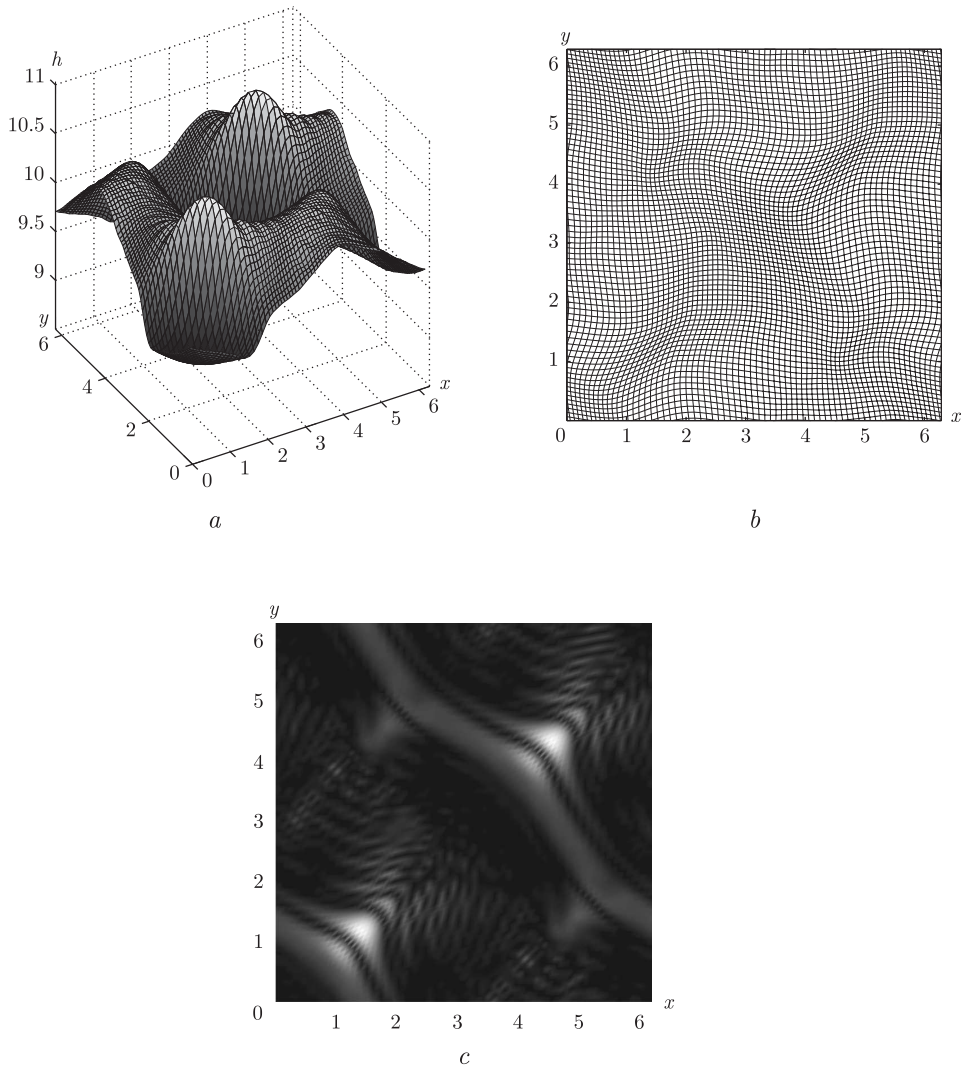}
\caption{Numerical integration of the two-dimensional shallow-water equations~\eqref{eq:TwoDimensionalShallowWater} using the scheme~\eqref{eq:2DShallowWaterCompCoordsScheme} with $\tau=0.001$ and $N_x\times N_y=71\times71$ grid points on the square $[0, 2\pi]\times [0, 2\pi]$ over the time interval $[0,2]$. The initial conditions are $u=A\sin(x+\varphi_0)\sin y$, $v=A\sin x\sin y$ and $h=h_0+A\cos(x+\varphi_0)\cos y$, with $A=0.4$, $\varphi_0=\pi/6$ and $h_0=10$. As a weight function, $w^2=\sqrt{1+\alpha(h_{xx}+h_{yy})^2}$ is chosen with $\alpha=0.4$.
({\it a}) Numerical solution for $h$ at $t=2$.
({\it b}) Spatial discretization grid at $t=2$.
({\it c}) The weight function at $t=2$.
}
\label{fig:EulerianImplicitInvariantSchemeTwoDimensionalShallowWaterResult}
\end{figure}

In Fig.~\ref{fig:EulerianImplicitInvariantSchemeTwoDimensionalShallowWaterResult} we repeat the numerical integration of the two-dimensional shallow-water equations with the setting of Fig.~\ref{fig:ImplicitInvariantSchemeTwoDimensionalShallowWaterResult} but now using the scheme~\eqref{eq:2DShallowWaterCompCoordsScheme} in combination with a grid generator based on~$w^2$.
Similarly to the case of the one-dimensional shallow-water equations, it can be seen from Fig.~\ref{fig:EulerianImplicitInvariantSchemeTwoDimensionalShallowWaterResult} that the usage of a grid generator leads to grids that are not as distorted as those obtained from a purely Lagrangian scheme. Moreover, the regions of grid concentration are now directly linked to the physical behavior of the numerical solution for the dependent variable~$h$. The scheme~\eqref{eq:2DShallowWaterCompCoordsScheme} is mass and momentum conserving but, as all the other schemes presented in the paper, dissipates the energy.
The conserved quantities are evaluated at time~$t$ as
$\mathcal M = \Delta\xi\Delta\eta\sum_{j,k}h_{jk}J_{jk}$,
$\mathcal P_x = \Delta\xi\Delta\eta\sum_{j,k}h_{jk}u_{jk}J_{jk}$,
$\mathcal P_y = \Delta\xi\Delta\eta\sum_{j,k}h_{jk}v_{jk}J_{jk}$ and
$\mathcal H = \frac12\Delta\xi\Delta\eta\sum_{j,k}(h_{jk}(u_{jk}^2+v_{jk}^2)+h_{jk}^2)J_{jk}$.

\section{Conclusion}\label{sec:Conclusion}

The present paper is devoted to the construction of several invariant numerical schemes modeling shallow-water dynamics. In particular, we aim to describe a possible bridge between the formalism of constructing invariant numerical schemes and the existing theory on adaptive moving meshes. Such a bridge was already indicated in the literature. Indeed, there already exist a number of investigations devoted to the importance of scale invariance in the theory of moving mesh equations. Thus, in~\cite{budd96Ay,budd09Ay,budd99Ay} (see also~\cite[p.\ 111]{huan10By} and references therein) a moving mesh partial differential equation was constructed that preserves the scale invariance of the physical differential equation to be discretized. The extension of this idea to setting up a grid generator that is invariant with respect to (a suitable subgroup of) the maximal Lie invariance group of a system of differential equations is therefore straightforward and was conceptually indicated in the aforementioned sources. The idea of introducing computational coordinates into invariant numerical schemes has also been successfully demonstrated for one-dimensional nonlinear Schr\"{o}dinger equations~\cite{budd01Ay}.

We require our discretization schemes to be invariant with respect to the subgroup of the maximal Lie symmetry group of the (resp.\ one- or two-dimensional) shallow-water equations admitted when imposing periodic boundary conditions. From the physical point of view it is natural to assume that appropriate symmetries of the system of differential equations under consideration act as equivalence transformations on a joint class of physically relevant boundary value problems. Imposing periodic boundary conditions for varying intervals in the one-dimensional case (resp.\ for rectangular domains of varying sizes whose sides are parallel to coordinate axes in the two-dimensional case) while both the initial time and initial conditions also vary, we select the subgroup generated by the time and space translations, the Galilean boosts and the scalings symmetries of the shallow-water equations. Other subgroups might be chosen as well, but for wide or even infinite-dimensional maximal Lie invariance (pseudo)groups admitted by the prominent models in hydrodynamics it can be quite intricate to justify the choice for such subgroups from the physical point of view.

\looseness=-1
In general, the inclusion of well-proven principles in the study of invariant numerical schemes is required. The invariant schemes for numerous evolution equations constructed so far were mostly purely Lagrangian schemes. However, these schemes are not in prevalent use in practice as they usually lead to complicated mesh geometries that might eventually (at least locally) degrade the quality of the numerical solution. This can be seen directly by comparing Figures~\ref{fig:ImplicitInvariantSchemeTwoDimensionalShallowWaterResult} and~\ref{fig:EulerianImplicitInvariantSchemeTwoDimensionalShallowWaterResult}, where the stronger distortion of the grid lines in the Lagrangian scheme is already manifest after a relatively short period of integration. Therefore, the formulation of invariant grid generators coupled with suitable invariant discrete counterparts of physical systems of differential equations is a practical way for symmetry preserving numerical integration of these systems.

A further novel feature of the present paper is the construction of invariant numerical schemes for higher-dimensional systems of partial differential equations. Higher-dimensional schemes are especially challenging if it is not possible to use fixed orthogonal grids. In the course of constructing such schemes for the two-dimensional shallow-water equations we have shown that invariant discretizations are not only restricted to finite difference schemes. It is possible and straightforward to also formulate finite volume discretizations that preserve symmetries of systems of differential equations. In a similar manner, other discretization techniques, such as the finite element method, could be employed as well.

This problem can also be tackled by transforming the system under consideration into computational coordinates. We have shown that the transition to these coordinates is a natural step in the course of the construction of Galilean invariant discrete schemes. The key to the construction is then not to simply combine difference invariants as proposed in the original method by Dorodnitsyn but to study the transformation laws of the equations in computational coordinates for the respective symmetries. These laws are trivial for all the admitted symmetries except for the Galilean boosts. For Galilean invariance it is found that the new momentum equations are given as the combination of the old momentum and continuity equations. An invariant discretization is therefore achieved by finding proper discrete counterparts of these transformation laws rather than combining difference invariants.

\looseness=-1
Because the main objective of this paper is to demonstrate different strategies for finding invariant discretization schemes exemplified with the shallow-water equations, the discretization schemes considered are kept as simple as possible. This concerns both the design of the schemes themselves (e.g., using only two-level time integration methods and unstaggered grids) and the solution of the resulting algebraic equations, which is done in the most direct and straightforward manner. More advanced integration and algebraic solution techniques can be readily adopted but their discussion is restrained to keep the focus of the paper on the conceptual aspect of introducing the Lie symmetry approach in the framework of numerical modeling as far as possible. For example, the extension to more advanced time integration methods such as arbitrary Runge--Kutta and general time-splitting schemes can be done by extension of the discretization stencils via inclusion of further time layers. Similarly, the usage of staggered grids can be facilitated by adding further points to the stencil on the same time layer. The procedure of invariant discretization involving wider stencils then follows precisely the same techniques as outlined and used in the present paper. Within the approach based on the construction of difference invariants, both ways of extending the stencils will lead to a larger number of invariants and thus to an increased number of possibilities for combining them to form a particular discretization scheme.

It was mentioned in the introduction that a system of differential equations might possess various qualitative properties that one should aim to preserve in the course of setting up a numerical model. Besides symmetries, it is of outstanding importance to monitor the behavior of conserved quantities possessed by the system under consideration. This is a problem of central importance in long-term integrations of such systems as a systematic failure in capturing conservation laws may lead to unrealistic numerical results (e.g., loss of mass or wrong turbulence spectra). Proper discretizations of the momentum form of the shallow-water equations conserve the mass and momenta exactly or to high order, but none of them is actually energy conserving. This should not come as a complete surprise as setting up energy-conserving schemes for the shallow-water equations is a quite nontrivial problem; see, e.g., the schemes proposed in~\cite{arak81Ay,salm07Ay,somm09Ay}. The inclusion of additional conserved quantities in the construction of invariant discretization schemes will therefore be one of our future research topics.

From a more general point of view, the requirement of preserving symmetries in a discretization scheme might lead to a geometric justification for using adaptive meshes. Though there are several classes of physical problems (such as blow-ups) for which adaptive meshes are well suited, the usage of such meshes is not undisputed in the numerical analysis and geophysical fluid dynamics communities. The drawbacks of moving meshes, such as an additional level of complexity of the schemes and the computational overhead resulting from computing and storing the mesh points at each time level, must be well opposed to their potential benefits on a case-by-case basis. The result that the numerical preservation of important structural properties like symmetries automatically requires using moving meshes can thus be seen as a geometric argument for allowing adaptive discretization grids for certain classes of physical differential equations. Moreover, the usage of a grid redistribution equation (or $r$-adaptivity) as advocated in the present paper is also most suitable because it can be efficiently implemented within the framework of parallel computing.

\subsection*{Acknowledgments}

The authors thank Prof.\ Pavel Winternitz and Dr.\ Matthias Sommer for useful discussions. The remarks by the anonymous referees are much appreciated. This research was supported by the Austrian Science Fund (FWF), projects J3182--N13 (AB), P20632 and Y237 (ROP).

{\footnotesize

}
\end{document}